\documentclass[12pt]{spieman}  % 12pt font required by SPIE;
\usepackage{amsmath,amsfonts,amssymb}
\usepackage{graphicx}
\usepackage{setspace}
\usepackage{tocloft}
\usepackage{multirow}
\usepackage[final]{pdfpages}
\usepackage{booktabs} % For \toprule, \midrule and \bottomrule
\usepackage{siunitx} % Formats the units and values
\usepackage{subcaption}
\usepackage{textcomp}

\usepackage{lineno}
\usepackage{hyperref}
\hypersetup{
    colorlinks=true,
    linkcolor=blue,
    filecolor=magenta,      
    urlcolor=cyan,
}

\title{WALOP-South: A Four-Camera One-Shot Imaging Polarimeter for \textsc{pasiphae}   Survey. Paper II - Polarimetric Modelling and Calibration}

\author[a*]{Siddharth Maharana}
\author[a,j]{Ramya M. Anche}
\author[a,b,e]{A. N. Ramaprakash}
\author[a]{Bhushan Joshi}
\author[g]{Artem Basyrov}
\author[b,c]{Dmitry Blinov}
\author[b,c]{Carolina Casadio}
\author[h]{Kishan Deka}
\author[g]{Hans Kristian Eriksen}
\author[h]{Tuhin Ghosh}
\author[g]{Eirik Gjerløw}
\author[b,c]{John A. Kypriotakis}
\author[b,c]{Sebastian Kiehlmann}
\author[b,c]{Nikolaos Mandarakas}
\author[f]{Georgia V. Panopoulou}
\author[b,c]{Katerina Papadaki}
\author[b,c]{Vasiliki Pavlidou}
\author[e]{Timothy J. Pearson}
\author[b,c]{Vincent Pelgrims}
\author[d,i]{Stephen B. Potter}
\author[e]{Anthony C. S. Readhead}
\author[b,c]{Raphael Skalidis}
\author[g]{Trygve Leithe Svalheim}
\author[b,c]{Konstantinos Tassis}
\author[g]{Ingunn K. Wehus}

\affil[a]{Inter-University Centre for Astronomy and Astrophysics, Post bag 4, Ganeshkhind, Pune, 411007, India}
\affil[b]{Institute of Astrophysics, Foundation for Research and Technology-Hellas, Voutes, 70013 Heraklion, Greece}

\affil[c]{Department of Physics, University of Crete, Voutes, 70013 Heraklion, Greece}

\affil[d]{South African Astronomical Observatory, PO Box 9, Observatory, 7935, Cape Town, South Africa}

\affil[e]{Cahill Center for Astronomy and Astrophysics, California Institute of Technology, Pasadena, CA, 91125, USA}

\affil[f]{Hubble Fellow, California Institute of Technology, Pasadena, CA 91125, USA}

\affil[g]{Institute of Theoretical Astrophysics, University of Oslo, P.O. Box 1029 Blindern, NO-0315 Oslo, Norway}

\affil[h]{School of Physical Sciences, National Institute of Science Education and Research, HBNI, Jatni 752050, Odisha, India}

\affil[i]{Department of Physics, University of Johannesburg, PO Box 524, Auckland Park 2006, South Africa}

\affil[j]{Steward Observatory, University of Arizona, Tucson, Arizona, 85721, USA}

\cftpagenumbersoff{figure}
\cftpagenumbersoff{table} 
\begin{document} 
\maketitle

\begin{spacing}{1}   % use double spacing for rest of manuscript

\begin{abstract}
The Wide-Area Linear Optical Polarimeter (WALOP)-South instrument is an upcoming wide-field and high-accuracy optical polarimeter to be used as a survey instrument for carrying out the Polar-Areas Stellar Imaging in Polarization High Accuracy Experiment (\textsc{pasiphae}  ) program. Designed to operate as a one-shot four-channel and four-camera imaging polarimeter, it will have a field of view of $35\times 35$ ~arcminutes and will measure the Stokes parameters $I$, $q$, and $u$ in a single exposure in the SDSS-r broadband filter. The design goal for the instrument is to achieve an overall polarimetric measurement accuracy of 0.1~\% over the entire field of view. We present here the complete polarimetric modeling of the instrument, characterizing the amount and sources of instrumental polarization. To accurately retrieve the real Stokes parameters of a source from the measured values, we have developed a calibration method for the instrument. Using this calibration method and simulated data, we demonstrate how to correct for instrumental polarization and obtain 0.1~\% accuracy in the degree of polarization, $p$. Additionally, we tested and validated the calibration method by implementing it on a table-top WALOP-like test-bed polarimeter in the laboratory.
\end{abstract}

% Include a list of up to six keywords after the abstract
\keywords{polarization, polarimetric modeling, polarimetric calibration, linear polarimetry, optical polarization, wide-field polarimeter, one-shot polarimetry}

% Include email contact information for the corresponding author
{\noindent \footnotesize\textbf{*}Siddharth Maharana,  \linkable{sidh@iucaa.in, sidh345@gmail.com} }

\section{Introduction}

Using the two upcoming Wide-Area Linear Optical Polarimeters\cite{walop_s_spie_2020} (WALOPs) as survey instruments, the Polar-Areas Stellar Imaging in Polarization High Accuracy Experiment (\textsc{pasiphae}) program aims to create the first large swath ($>$~4000 square degrees) optopolarimetric map of the sky, towards the Galactic polar regions. The main objectives of the \textsc{pasiphae}   program include - (i) determining the 3D structure of the dust distribution along a large number of lines of sight with the sub-degree plane of sky angular resolution, (ii) determining the plane of the sky orientation of the magnetic fields associated with the multiple dust clouds that are generally seen along each line of sight\cite{Panopoulou_2019, vincent_decorrelation_paper, vincent_tomography, gina_lenz}, (iii) to test the physics of interstellar dust, especially concerning grain alignment and size distribution, and (iv) to trace paths traversed by ultra-high-energy cosmic rays through  the  Galaxy. For an extensive description of the motivation and scientific objectives of the \textsc{pasiphae} program, we refer the reader to the \textsc{pasiphae} white paper by Tassis et al.\cite{tassis2018pasiphae}. The survey will be concurrently executed from the southern and northern hemispheres, using the WALOP-South instrument mounted on the 1~m telescope at South African Astronomical Observatory’s (SAAO) Sutherland Observatory and WALOP-North mounted on the 1.3~m telescope at Skinakas Observatory, Greece, respectively. Table~\ref{techtable} captures the design goals of the WALOP instruments, which were decided based on the scientific goals of  the \textsc{pasiphae} program as well as the current status of state-of-the-art polarimetric instrumentation capabilities. This has been discussed in the optical design paper of the WALOP-South instrument by Maharana et al.\cite{WALOP_South_Optical_Design_Paper}, hereafter to be referred to as Paper I.

\begin{table}
    \centering
    \begin{tabular}{ccc}
        \hline
        \textbf{Sl. No}. & \textbf{Parameter} & \textbf{Technical Goal} \\
         \hline
        %1 & Polarimetric Sensitivity ($s$) & 0.05~\%\\
        % \hline
        1 & Polarimetric Accuracy ($a$) & 0.1~\%\\
         %\hline
        2 & Polarimeter Type & Four Channel One-Shot Linear Polarimetry \\
         %\hline
        3 & Number of Cameras & 4 (One Camera for Each Arm)\\
         %\hline
        4 & Field of View & $30\times30$~arcminutes\\
         %\hline
        5 & Detector Size & $4k\times4k$ (Pixel Size = $15~{\mu}m$) \\
        %\hline
        %9 & Pixel Size & $15~{\mu}m$\\
        %\hline
        6 & No. of Detectors & 4 \\
        %\hline
        7 & Primary Filter & SDSS-r \\
        %\hline
        8 & Imaging Performance & Close to seeing limited PSF \\
         \hline
    \end{tabular}
    \caption{Design goals for WALOP instruments.}
    \label{techtable}
\end{table}

Both WALOPs are currently under development at the Inter-University Centre for Astronomy and Astrophysics (IUCAA), Pune, India. Of the two WALOP instruments, WALOP-South is scheduled first for commissioning in 2022. Paper I gives a complete description of the optical model of the WALOP-South instrument. The optical model of WALOP-North is very similar to WALOP-South- the differences are owing to the differences in the telescope optics as well as optomechanical constraints. Consequently, the polarimetric behavior of the two instruments is qualitatively identical. In this paper, we focus on WALOP-South to illustrate the polarimetric modeling and calibration approach and its validation.

Integrated (and unresolved) light from most stars is unpolarized in optical wavelengths. However, as light from the stars passes through the intervening interstellar medium, dichroic extinction in the line of sight and anisotropic scattering by elongated grains aligned with the magnetic field in dust clouds introduces linear polarization ($p$) to fractions of around a few percent or less, depending on the dust column density and the geometry of the magnetic field\cite{andersson_review, Hensley_2021, Heiles_2000}. As the \textsc{pasiphae}  survey is aiming to cover the Galactic polar regions where the interstellar dust is very sparse\cite{Skalidis}, the expected amount of polarization is even smaller ($<$~ 0.3 \%). We define polarimetric sensitivity ($s$) as the least value and change of linear polarization that the instrument can measure, without correction for the instrumental polarization. $s$ is a measure of the internal noise and random systematics of the instrument. Polarimetric accuracy ($a$) is the measure of closeness of the predicted polarization of a source to the real value after correcting for the instrument induced polarization using calibration techniques. The technical goal for the WALOP instruments is to obtain polarimetric sensitivity ($s$) and accuracy ($a$) of 0.05~\% and 0.1~\%, respectively. To determine the intrinsic (on-sky) linear Stokes parameters of a source, $q$ and $u$, from the instrument measured Stokes parameters $q_{m}$ and $u_{m}$, polarimetric calibrations are required to estimate and correct for instrument induced polarimetric effects.% such as zero-offsets and cross-talk.

While polarimeters in astronomy achieving polarimetric sensitivity and accuracy of around $10^{-3}~\%$ in $p$ have been made, eg. HIPPI-2 \cite{Bailey_2020} and DIPOL-2 \cite{DIPOL2}, they have very narrow fields of view (FoV), effectively only capable of point source polarimetry. In narrow FoV polarimeters, the rays are incident at nearly normal and/or azimuthally symmetric angles on all the optical elements, leading to very small levels of net instrumental polarization. Whereas, in the case of wide-field polarimeters like WALOPs, rays from off-axis field points propagate inside the optical system at oblique angles. As these angles become larger, the polarization effects due to the optics become stronger. All optical elements can introduce instrumental polarization in the following ways: (i) oblique angles of incidence on an optical surface lead to the preferential transmission of one orthogonal polarization over the other- this can be described through Fresnel coefficients, and (ii) retardance and consequent polarimetric cross-talk due to stress birefringence, as a result of thermal and mechanical stresses on the optics. Over and above these, the main source of instrumental polarization in WALOPs arises from the angle of incidence-dependent retardance by the half-wave retarder plates (HWP) in the Wollaston Prism Assembly (WPA) of the instruments. To estimate and correct for these effects so as to accurately recover the real Stokes parameters, we have developed a detailed method for calibration of the instruments. Temperature change and instrument flexure are known sources of variable instrumental polarization \cite{Tinyanont2018}. The optomechanical design of the instrument (Maharana et al., in preparation) has been created to ensure the time-independent optical and polarimetric behavior of the instrument. For example, irrespective of the variation of these two factors, the optics holders and overall instrument model have been designed to (a) maintain the overall instrument optical alignment to tight tolerance levels and (b) minimize the stress birefringence on the lenses to levels below the instrument sensitivity (Anche et al., in preparation). %We find that the theoretical model and experimental results obtained on this system for the calibration method are consistent with each other.} 

\par In this paper, we present the polarization modeling and calibration method for the WALOP-South instrument, which are presented in Sections~\ref{WALOP_modelling_section} and \ref{Calibration_model_theortical}, respectively. The instrument polarization model was developed using the instrument's optical model with the aid of polarization analysis features of the \href{https://www.zemax.com/}{Zemax}\textsuperscript{\textregistered} software, which was used previously in polarization modeling work for other telescopes and instruments such as the Thirty Meter Telescope by Anche et al.\cite{Ramya_TMT_polarization} and Daniel K. Inouye Solar Telescope by Harrington et al.\cite{dkist_polarization_modelling}. We have tested and validated the calibration method on the computer-based Zemax optical model of the WALOP-South instrument and have achieved better than 0.1~\% accuracy across most of the FoV. In this model, various on-sky effects such as photon noise, the variable transmission of sky as well as non-idealities in calibration optics were incorporated. Further, to test the calibration model on real WALOP-like polarimeter systems, a table-top test-bed polarimeter system was developed and assembled in the lab at IUCAA. Section~\ref{calibration_model_lab_test} contains details of the lab set-up and results obtained from it. Finally, Section~\ref{calibration_conclusion} contains conclusions and further discussions.

\section{Polarimetric Modelling of WALOP-South}\label{WALOP_modelling_section}

\subsection{Optical Model of the WALOP-South Instrument}

\begin{figure}
    \centering
    \includegraphics[width=1\linewidth]{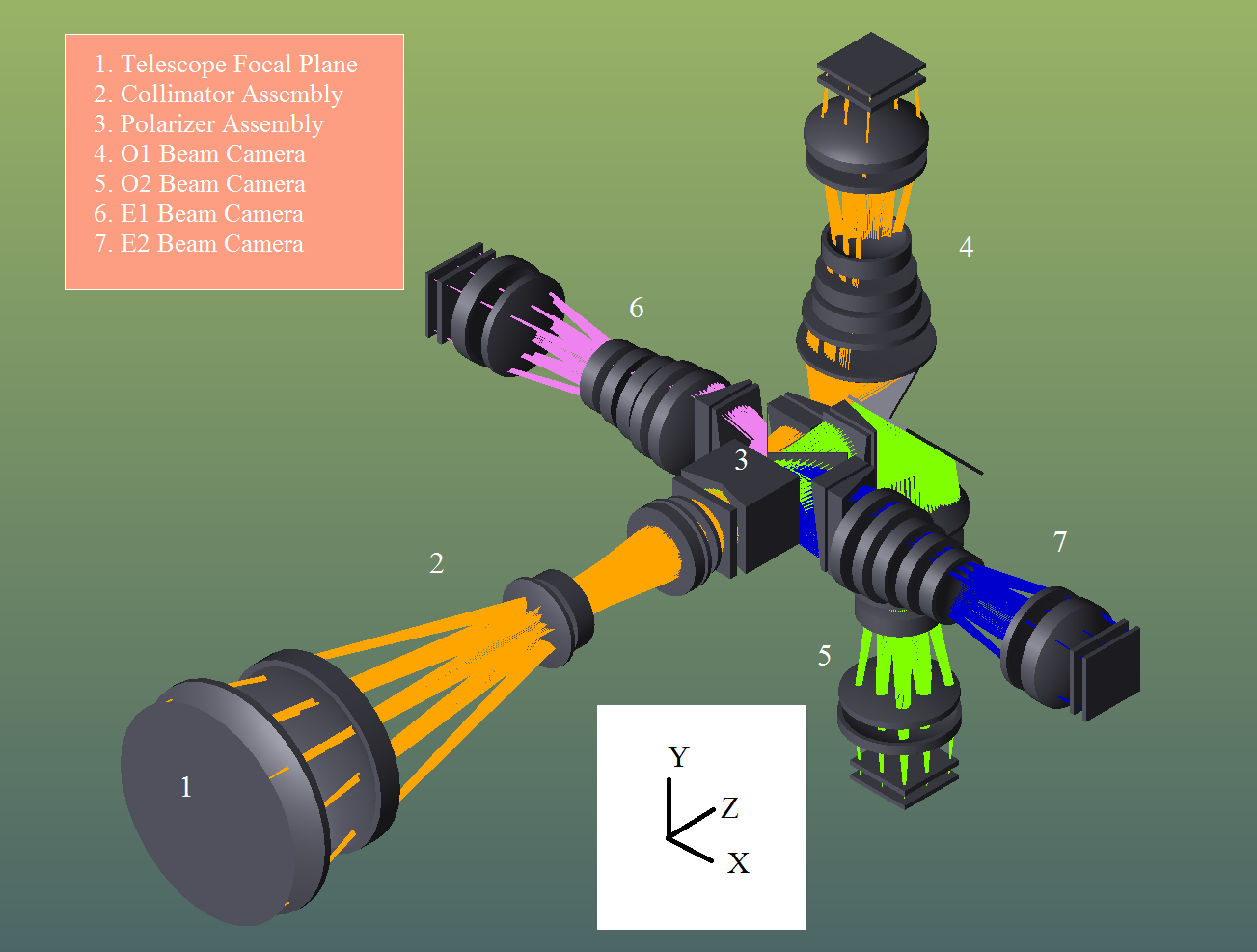}
    \caption{Optical layout of the WALOP-South instrument (Paper I). Starting at the telescope focal plane, it accepts the beam for the entire field and  creates a pupil through the collimator assembly. Located at the pupil, the polarizer assembly splits the beam into four channels corresponding to $0^{\circ}$, $45^{\circ}$, $90^{\circ}$ and $135^{\circ}$ polarization angles, referred to as O1, O2, E1 and E2 beams, respectively. Further, the polarizer assembly folds and steers the O beams along the +y and -y directions and the E beams along the +x and -x directions. Thereon, each beam is imaged on a $4k\times4k$ CCD detector through its own camera.}
    \label{WALOP-S_shaded_model}
\end{figure}

Here we present an overview of the optical model of the WALOP-South instrument; we refer the reader to Paper I for a detailed description. Fig~\ref{WALOP-S_shaded_model} shows the 3D optical layout of the instrument. The optical system consists of the following assemblies: a collimator, a polarizer assembly, and four cameras (one for each channel). The collimator assembly begins from the telescope focal plane. Aligned along the z-axis, it creates a pupil image that is fed to the polarizer assembly. The polarizer assembly acts as the polarization analyzer system of the instrument and splits the pupil beam into four channels corresponding to $0^{\circ}$, $45^{\circ}$, $90^{\circ}$ and $135^{\circ}$ polarization angles, which are referred to as O1, O2, E1, and E2 beams, respectively. Additionally, this assembly folds and steers the O beams along the +y and -y directions and the E beams along the +x and -x directions. Each channel has its own camera to image the entire FoV on a $4k\times4k$ CCD detector. The obtained FoV of the instrument is $34.8\times34.8$~arcminutes, surpassing the design goal of $30\times30$~arcminutes. Table~\ref{op_design_summary} lists the key design parameters of the instrument's optical system. %The polarizer assembly is the most novel and complex aspect of the WALOP-South optical design, and Section~\ref{pol_ass} describes it's architecture and working. As part of the optical design, we also designed a guider camera for instrument as well as new baffles for the telescope to accommodate the large field of view of WALOP-South- these are described in Paper I.

\iffalse
The complete optical design of the instrument is described in elaborate detail in the detail in Paper I. Fig~\ref{WALOP-S} shows the optical model of the WALOP-South instrument, designed in the Zemax software. The instrument's optics begins at the telescope focal plane where the collimator is the first optical sub-assembly, followed by the polarizer assembly which analyzes the light, and then there are four camera assemblies that image the four channels into four detectors. Fig~\ref{wpa_cartoon} shows the working of the polarizer assembly, and how the polarization state of the light is getting changed.
\fi

\begin{table}[htbp!]
    \centering
    \begin{tabular}{cc}
        \hline
        \textbf{Parameter} & \textbf{Design Value/Choice}\\
        \hline
        Filter & SDSS-r \\
        %\hline
        Telescope f/number & 16.0 \\
        %\hline
        Camera f/number & 6.1 \\
        %\hline
        Collimator Length & 700~mm\\
        %\hline
        Camera Length & 340~mm \\
        %\hline
        No of lenses in Collimator & 6 \\
        %\hline
        No of lenses in Each Camera & 7 \\
        %\hline
        Detector Size & $4096\times4096$, $15~{\mu}m$ pixel. \\
        %\hline
        Plate scale at detector & 0.51"/pixel\\
        \hline
        
    \end{tabular}
    \caption{Values of the key parameters of WALOP-South optical design.}
    \label{op_design_summary}
\end{table}

%\subsection{Polarizer Assembly Design}\label{pol_ass} 
\begin{figure}
    \centering
    \includegraphics[width=1\linewidth]{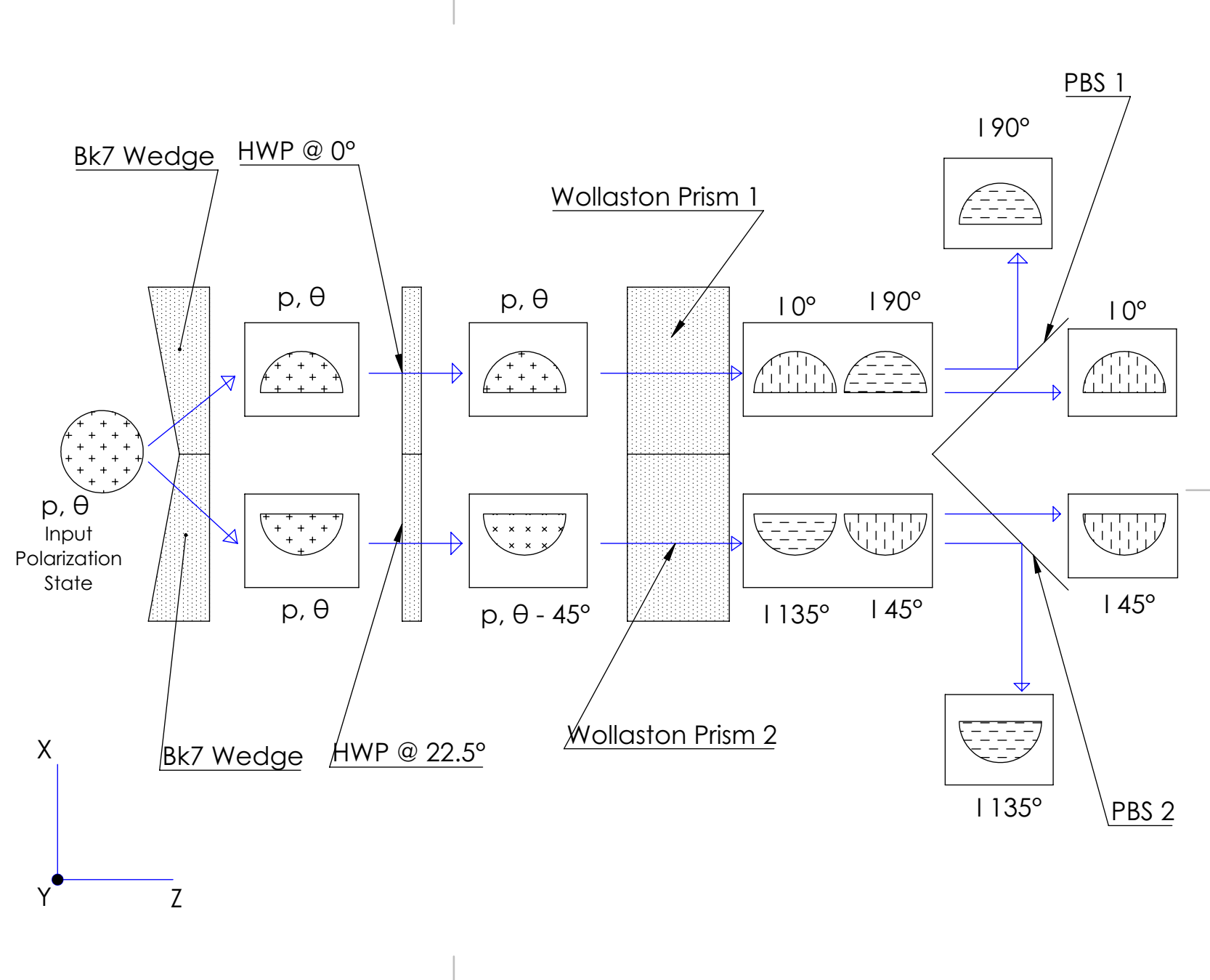}
    \caption{A cartoon illustration of the working of the polarizer assembly of the WALOP-South instrument (Paper I). Together, the Wollaston Prism Assembly consisting of the two BK7 glass wedges, Wollaston Prisms (WP) and Half-Wave Plates (HWP) and two Polarization Beam-Splitter (PBS) act as the polarization beamsplitter sub-system. The pupil is split between the two BK7 wedges which is then fed to the twin HWP + WP system to be split into four channels with the polarization states of $0^{\circ}$, $45^{\circ}$, $90^{\circ}$ and $135^{\circ}$. Afterwards, the two PBS' direct these four beams in four directions. $p$ and $\theta$ shown are as seen in the x-y plane when viewed along the z-axis of the cartoon and the change in the polarization state of the beams while passing through this system is annotated. }
    \label{pol_ass_cartoon}
\end{figure}

The Polarizer Assembly consists of four sub-assemblies: (i) Wollaston Prism Assembly (WPA), (ii) Wire-Grid Polarization Beam-Splitter (PBS), (iii) Dispersion Corrector Prisms (DC Prisms), and (iv) Fold Mirrors. The WPA consists of two identical calcite Wollaston Prisms (WP), with a half-wave plate (HWP) and a BK7 glass wedge in front of each WP (Fig~\ref{pol_ass_cartoon}). The WPs have an aperture of $45\times80~mm$ and a wedge angle of $30^{\circ}$, resulting in a split angle of $11.4^{\circ}$ at $0.6~{\mu}m$ wavelength. The left WP has a HWP with fast-axis at $0^{\circ}$ with respect to the instrument coordinate system (ICS) to separate the $0^{\circ}$ and $90^{\circ}$ polarizations while the right WP has a HWP with fast-axis at $22.5^{\circ}$ to split the $45^{\circ}$ and $135^{\circ}$ polarizations. The BK7 wedges at the beginning of the WPA, which share the incoming pupil beam equally, ensure that rays from off-axis objects in the FoV entering at oblique angles of incidence do not hit the interface between the WPs, which will lead to throughput loss as well as instrumental polarization from scattering arising at the surface. The PBS' act as beam selectors, allowing both the O1 and O2  beams to pass through while folding the E1 and E2 beams along -x and +x directions. Fig~\ref{pol_ass_cartoon} shows the overall working idea of the WPA and PBS components of the polarizer assembly. The Dispersion Corrector (DC) Prisms are a pair of glass prisms present in the path of each of the four beams after the PBS' to correct for the spectral dispersion introduced by the WPA (Paper I). Additionally, mirrors placed at $\pm~45^{\circ}$ the y-z plane fold the O beams into +y and -y directions to limit the total length of the instrument to 1.1~m from the telescope focal plane.

\subsection{Mueller Matrix Modelling}
The entire instrument's polarization behavior can be mathematically modeled as an Instrument Matrix ($M_{inst}$), similar to a Mueller matrix, as shown in Equation~\ref{mueller_matrix_instrument}. A brief description of $M_{inst}$ and its relation to the Mueller matrices of the instrument's four-camera optical system is presented in Appendix~\ref{instrument_matrix_appendix}. As we are interested in the normalized Stokes vectors, it can be rewritten in the form of Equation~\ref{mueller_matrix_instrument_normalized}. The first row of $m_{inst}$ has been ignored as it does not affect the measured Stokes parameters. The $m_{22}$, $m_{33}$ and $m_{44}$ terms are the \textit{polarimetric efficiencies} of the instrument which capture how each of the measured Stokes parameters scale with respect to their corresponding input values. The $m_{21}$, $m_{31}$ and $m_{41}$ terms are referred to as the polarimetric \textit{zero offsets} as they represent the measured Stokes parameters when the input is unpolarized. The terms $m_{23}$ and $m_{32}$ capture the instrument \textit{cross-talk} between linear Stokes parameters, i.e., how much of $u$ is converted into $q_{m}$, and $q$ into $u_{m}$, respectively. Likewise, the terms $m_{24}$ and $m_{34}$, and $m_{42}$ and $m_{43}$ capture the \textit{cross-talk} between the circular and linear polarizations. For an ideal polarimeter, the diagonal terms will be 1 and all other terms will be zero. For linear polarimeters, the fourth row concerned with $v_{m}$ is ignored. For most celestial objects, $v$ is usually at least an order of magnitude smaller than $q$ and $u$ at optical wavelengths and hence the fourth column can be ignored for linear polarimeters. %, as shown in Equation~\ref{mmatrix_instrument_ideal_intro}.
% $M_{42}$ and $M_{43}$ capture the cross-talk from the linear polarization to measured circular polarization. 

\begin{gather}\label{mueller_matrix_instrument}
S_{m} = I_{m} \times
\begin{bmatrix}
1 \\
q_{m} \\
u_{m} \\
v_{m} 
\end{bmatrix}
= M_{inst}\times S = 
\begin{bmatrix}
M_{11} & M_{12} & M_{13} & M_{14}  \\
M_{21} & M_{22} & M_{23} & M_{24}  \\
M_{31} & M_{32} & M_{33} & M_{34}  \\
M_{41} & M_{42} & M_{43} & M_{44} 
\end{bmatrix}
\times  I 
\begin{bmatrix}
1 \\
q \\
u \\
v 
\end{bmatrix}
\end{gather}

\begin{gather}
s_{m} = 
\begin{bmatrix}
1 \\
q_{m} \\
u_{m} \\
v_{m} 
\end{bmatrix}
= 
m_{inst}\times s
= 
\begin{bmatrix}
m_{11} & m_{12} & m_{13} & m_{14}  \\
m_{21} & m_{22} & m_{23} & m_{24}  \\
m_{31} & m_{32} & m_{33} & m_{34}  \\
m_{41} & m_{42} & m_{43} & m_{44} 
\end{bmatrix}
\times  
\begin{bmatrix}
1 \\
q \\
u \\
v 
\end{bmatrix} \nonumber
\\
=
\begin{bmatrix}
- & - & - & -  \\
1 \,\to\,q_{m} & q\,\to\,q_{m} & u\,\to\,q_{m} & v\,\to\,q_{m}  \\
1 \,\to\,u_{m} & q\,\to\,u_{m} & u\,\to\,u_{m} & v\,\to\,u_{m}  \\
1 \,\to\,v_{m} & q\,\to\,v_{m} & u\,\to\,v_{m} & v\,\to\,v_{m}
\end{bmatrix}
\times  
\begin{bmatrix}
1 \\
q \\
u \\
v 
\end{bmatrix}
\label{mueller_matrix_instrument_normalized}
\end{gather}

%For an ideal linear polarimeter, the instrumental Mueller Matrix will look like .

%\textbf{Sources of instrumental polarization and cross-talk:}
\par For polarimeters with narrow FoV (few arcminutes or less), e.g. RoboPol\cite{robopol} and Impol\cite{impol}, the values of cross-talk terms are $\simeq$~0 and the polarimetric efficiency is $\simeq$~1. The main sources of instrumental polarimetric effect are the polarimetric zero offset terms $m_{21}$ and $m_{31}$. On sky, both these terms can be found by observing zero and partially-polarized standard stars.

For polarimeters with non-negligible cross-talk terms, a more detailed calibration method is needed. A major source of cross-talk is the presence of mirror systems in the optical path before the polarization analyzer system, as is often the case with instruments mounted on side-ports of telescopes or on a Nasmyth focus. For such instruments, an accurate Instrument matrix can be created by placing a calibration linear polarizer before any polarimetric effects are introduced (usually at the beginning of the instrument) \cite{instrumentation_book_springer,IRDIS_Calibration}.

\subsection{Polarimetric Cross-talk and Zero-offset}\label{WALOP_modelling_section_overall}
The measured Stokes parameters can be written in the form of Equations~\ref{calibration_eqn_q} and \ref{calibration_eqn_u}. In general, these can depend on all the intrinsic Stokes parameters of the source. Additionally, the coefficients in these equations are a function of the field position. To estimate the coefficients, the instrument can be made to observe sources with known states of polarization. 

\begin{equation}\label{calibration_eqn_q}
    q_{m} = m_{21} + m_{22}q + m_{23}u  
\end{equation}

\begin{equation}\label{calibration_eqn_u}
    u_{m} = m_{31} + m_{32}q + m_{33}u    
\end{equation}

\begin{figure}
\centering
\includegraphics[width=1\linewidth]{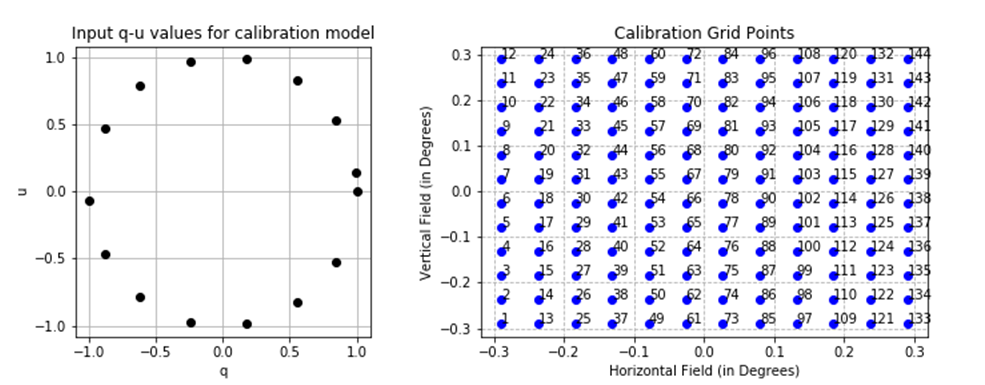}
\caption{Left: States of input polarized light given to WALOP-South Zemax model for creating the instrument calibration model, Right: grid points across WALOP-South FoV of $0.58^{\circ}\times0.58^{\circ}$ used for creating the instrument's polarimetric and calibration models.}
\label{grid_points_pol_input}
\end{figure}
%\begin{figure}
%\begin{subfigure}{0.49\textwidth}
    %\centering
    %\frame{\includegraphics[scale = 0.5]{input_qu.png}}
   % \caption{}
   % \label{input_qu}
%\end{subfigure}
%\begin{subfigure}{0.49\textwidth}
   % \centering
   % \frame{\includegraphics[scale = 0.5]{Grid_Plot_144.png}}
   % \caption{}
   % \label{Calibration Model Grid Points}
%\end{subfigure}

%\caption{ \textcolor{blue}{(a) States of input polarized light given to WALOP-South Zemax model for creating the instrument calibration model, and (b) grid points across WALOP-South FoV of $0.58^{\circ}\times0.58^{\circ}$ used for creating the instrument's polarimetric and calibration models.}}%. Input polarization states used for creating the calibration model grid points over WALOP-South FoV for which the calibration model was made.

We used the \textit{polarization transmission} tool in \href{https://www.zemax.com/}{Zemax}\textsuperscript{\textregistered} software to find the transmission of the optical system (here telescope + WALOP-South instrument) for each of the four detectors as a function of intrinsic source polarization and field position. For this, the instrument's FoV was divided into a $12\times12$ square grid (Fig~\ref{grid_points_pol_input}). Since WALOP-South will be working in the SDSS-r band, we obtain the cumulative transmission for the entire filter. We wrote a Zemax Programming Language script to give as input the list of grid points and input polarizations for which the transmission was obtained on all the four detectors (for more details, we refer the reader to the Zemax manual). We refer to the intensities of the beams along the $0^{\circ}$, $90^{\circ}$, $45^{\circ}$ and $135^{\circ}$ polarizations (O1, E1, O2 and E2, respectively) as $N_{0}$, $N_{1}$, $N_{2}$ and $N_{3}$. The measured Stokes parameters are then given by the following equations:

 %used previously in polarization modelling work for other telescopes and instruments such as the Thirty Meter Telescope by Anche et al. \cite{Ramya_TMT_polarization}. This analysis feature enables us

\begin{equation}\label{q_definition_cal}
 q_m = \frac{ N_{0} - N_{1}}{ N_{0} + N_{1}}
\end{equation}

\begin{equation}\label{u_definition_cal}
 u_m = \frac{ N_{2} - N_{3}}{ N_{2} + N_{3}}
\end{equation}

 % will assume that from now on.

\begin{figure}
\centering
\includegraphics[width=1\linewidth]{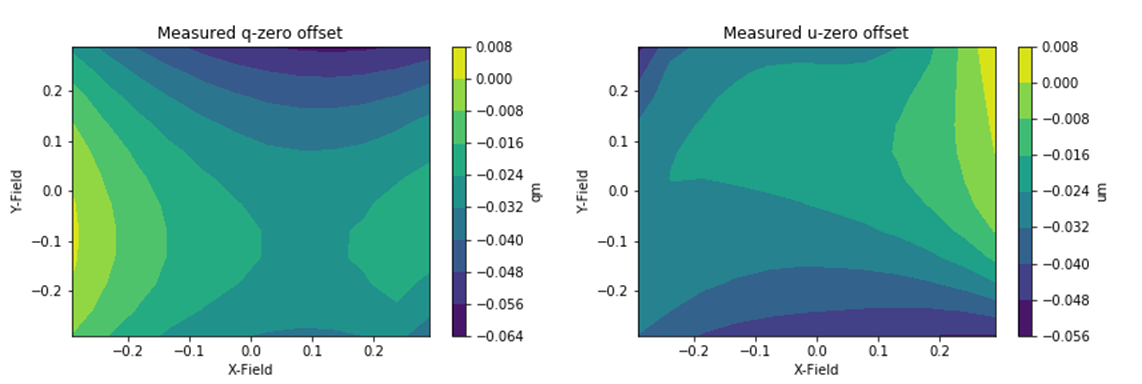}
\caption{WALOP-South instrument's zero-offset maps for $q_m$ (left) and $u_m$ (right). Polarization values are expressed in fractional units. The x and y field coordinates are in degrees.}
\label{WALOP-South Instrument Zero Polarization Map}
\end{figure} 

\par Fig~\ref{WALOP-South Instrument Zero Polarization Map} plots the maps of polarimetric zero offsets for $q_{m}$ and $u_{m}$ measurements, i.e, these are the measured Stokes parameters when the input to the system is unpolarized light. As expected, these resemble hyperbolic functions  since the differential Fresnel coefficients for orthogonal polarizations for curved surfaces such as lenses lead to similar spatial variation across the FoV \cite{moonlight_calibration_VLT}. The zero-offsets are seen to be as large as few percents in some parts of the field. Such relatively large values are expected as the E and O beams follow different optical paths (Fig~\ref{WALOP-S_shaded_model}). In particular, (i) unlike E-beams, the O-beams have a fold mirror in the path just after the PBS, and (ii) the transmission of the PBS for the O-beams is relatively inferior ($\sim90 ~\%$) compared to the near 100~\% reflectivity for the E beams.%has different t

\par
\begin{figure}
\centering
\includegraphics[width = 1\linewidth]{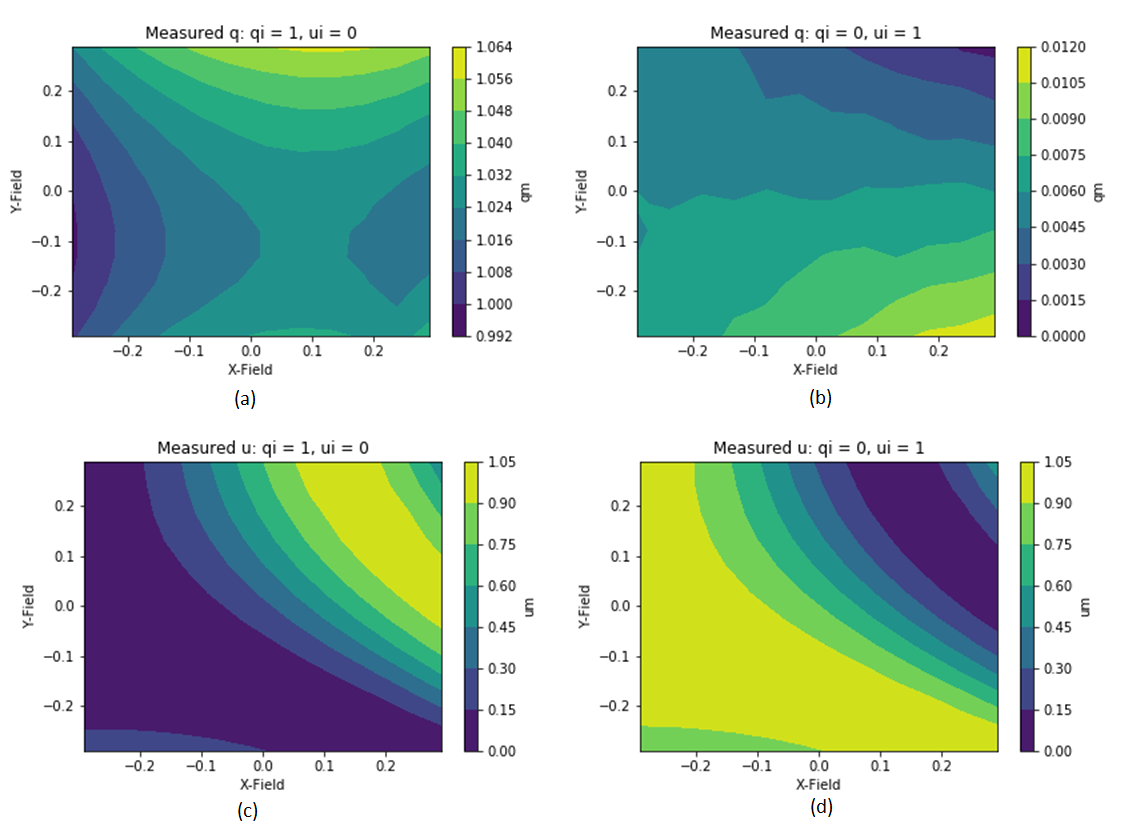}
\caption{Cross-talk (b and c) and polarimetric efficiency (a and d) maps for $q_m$  and $u_m$  in WALOP-South instrument. The x and y field coordinates are in degrees.}
\label{WALOP-South Instrument Polarization}
\end{figure}

 Fig~\ref{WALOP-South Instrument Polarization} shows the polarimetric efficiency and cross-talk maps for the $q_{m}$ and $u_{m}$ measurements. Figs~\ref{WALOP-South Instrument Polarization}~(a) and (c) are the measured Stokes parameter maps (zero offset corrected) when the input polarization is $q = 1$, i.e., these are the maps of $m_{22}$ and $m_{32}$ terms. For an ideal polarimeter, these should be 1 and 0 across the FoV. Figs~\ref{WALOP-South Instrument Polarization}~(b) and (d) are the measured Stokes parameter maps (zero offset corrected) when the input polarization is $u = 1$, so these are maps of the $m_{23}$ and $m_{33}$ terms. For an ideal polarimeter, these should be 0 and 1 across the FoV. % Likewise, Figs~\ref{qm_plot2} and \ref{um_plot2} are the measured Stokes parameter maps when the input polarization is $q = -1$, which should be -1 and 0 across the FoV.  Likewise, Figs~\ref{qm_plot4} and \ref{um_plot4} are the measured Stokes parameter maps when the input polarization is $u = -1$, which should be 0 and -1 across the FoV. 

\par As can be inferred from the plots, while the $m_{22}$ and $m_{23}$ terms are nearly close to their ideal values of 1 and 0 across the whole field, the same is not true for the $m_{32}$ and $m_{33}$ terms. The value of both $m_{32}$ and $m_{33}$ terms is well behaved and near to 0 and 1, respectively, in the lower-left half of the FoV, but it starts deviating rapidly in the upper-right half. In fact, in some places the $m_{32}$ tends to be 1 and $m_{33}$  tends to 0, i.e., all of the $u_{m}$ is actually derived from $q$ with very little contribution from $u$. Consequently, we find that while $q_{m}$ is mostly dependent only on $q$ and not so much on $u$, $u_m$ has strong dependence on $q$ and $u$-  i.e., the cross-talk terms are very significant for $u_m$. This dependence changes significantly across the FoV. Hence, we need to develop a calibration model that can create accurate mapping functions between the measured and real Stokes parameters across the FoV. The detailed working of this model is presented in Section~\ref{Calibration_model_theortical}. 

Due to the introduction of the two BK7 wedges just before the HWPs in the WPA (Fig~\ref{pol_ass_cartoon}), rays from the entire field enter at obliques angles of incidence to the HWPs. This leads to non-half wave retardance in the outgoing beam. In addition to this, due to manufacturing difficulties of the relatively large aperture HWPs that are needed, first-order HWPs have been used instead of zero-order HWPs. First-order HWPs are much more sensitive to oblique angles of incidence and have sharper deviation from half-wave retardance. Appendix~\ref{HWP_modelling_section} contains the analytical polarimetric modeling of the HWPs in the WPA based on ray tracing programs written in Python, which are the source of the observed cross-talk in the instrument. %The predominant source of cross-talk in $u_m$ is the non-half wave retardance of the HWPs used in the WPA and can be understood by modelling the Mueller matrix of the HWPs. This is presented in detail in the next section.%So while the measurements $q_{m}$ is not affected by cross-talk is independent of $u$, $u_{m}$ can have significant dependence on $q$. 

%Correspondingly, discussion on the plots. add a para. some observations. say we can undertsand this in terms of the reatrdance of the HWP. etc.

%\subsection{Polarimetric Modelling of WALOP HWPs} 

\section{Theoretical Calibration Model of WALOP-South}\label{Calibration_model_theortical}

% \textcolor{blue}{As part of the calibration process, a linear polarizer mounted on a motorized rotation stage is placed at the beginning of the instrument (after the telescope optics and before the first collimator lens). Through a motorized linear motion system, this stage will be inserted in the optical path during calibration observations, and removed during the main science observations. This calibration linear polarizer (CLP) will be used to provide as input linearly polarized light with different Electric Vector Polarization Angles (EVPA) to the instrument (i.e., different $q-u$ values), and will be used to create a mapping function between the instrument measured and real Stokes parameters of a source. Additionally, the model uses  polarized and unpolarized sources on sky for building and testing the of the model. While the model has been developed and tested on the WALOP-South instrument, respectively it will be valid for WALOP-North as well since their optical designs are very similar.} Content from this para will be used to fill in any info missing from the succeeding para.

Astronomers, both night-time as well as solar, in the past have calibrated many high polarization cross-talk instruments \cite{IRDIS_Calibration,Harrington_solar_Calibration}. The standard procedure involves creating a mapping function between instrument measured and real Stokes parameters of a source, often in the form of a matrix. For this, known polarized sources are needed to find the coefficients such as in Equations~\ref{calibration_eqn_qi} and \ref{calibration_eqn_ui}. We define $q_{i}$ and $u_{i}$ as the predicted Stokes parameters after applying the calibration model correction to the measured Stokes parameters. Due to the scarcity of on-sky calibration sources, a calibration linear polarizer (CLP) has been provided in the instrument before any polarization inducing optics, including the lenses. Mounted on a motorized rotary stage, the CLP will be inserted in the optical path during calibration observations, and removed during the main science observations. The CLP will provide as input linearly polarized light with different Electric Vector Position Angles (EVPA) to the instrument (i.e., $p \sim  1$ with different $q-u$ values). It is well known that the polarization introduced by a Cassegrain telescope optics system at its direct focus is at least an order of magnitude smaller than our target accuracy levels of 0.1~\%, and can be assumed as a non-polarizing component\cite{Sen_telescope_polarization}. Below we provide a prescription to develop a calibration model for WALOP-South which allows us to determine the coefficients of the polynomial equations of the form given in Equations~\ref{calibration_eqn_qi} and \ref{calibration_eqn_ui}. We find that a second order polynomial, without the ${q_m}{u_m}$ cross-term, provides a better fit to data and consequently gives a better polarimetric accuracy than linear equations. %The inadequacy of linear equations to provide the best fitting for calibration equations suggests that Mueller matrix operations are unable to capture many aspects to polarization in WALOP-South which lead to second order effects.
\begin{equation}\label{calibration_eqn_qi}
    q_{i} = a_{1} + b_{1}q_{m} + c_{1}u_{m} + d_{1}q_{m}^{2} + e_{1}u_{m}^{2}
\end{equation}

\begin{equation}\label{calibration_eqn_ui}
    u_{i} = a_{2} + b_{2}q_{m} + c_{2}u_{m} + d_{2}q_{m}^{2} + e_{2}u_{m}^{2}
\end{equation}

\subsection{Calibration model}

The WALOP-South FoV is divided into a grid of 12$\times$12 = 144 points, as shown in Fig~\ref{grid_points_pol_input}. For each grid point, the calibration model is developed independently. For this work, the \textit{polarization transmission} tool of the Zemax software was used to find the transmission at each of the four detectors as a function of polarization. As input, (i) unpolarized, and (ii) 15 states of 100\% linearly polarized beams with different EVPA ($\theta$) of same intensity were fed to the telescope+instrument, as shown in Fig~\ref{grid_points_pol_input} and the corresponding transmissions at the four detectors were obtained. Further, during processing of the data, we introduce the following effects on all the obtained transmissions: (i) random sky transparency variation between 0.7 to 1, (ii) assuming the star and sky magnitude to be 12 magnitude and 20 magnitude per arcseconds squared in the R band, we add the expected photometry noise (star + sky) to these transmissions, and (iii) effect of non-ideal nature of the CLP (in the WALOP instruments, we are using a Thorlabs polarizer whose extinction ratio is $>~5\times10^{3}$ in the R band: \url{https://www.thorlabs.com/thorproduct.cfm?partnumber=LPVISE2X2}). 

\begin{figure}
\includegraphics[width = 1\linewidth]{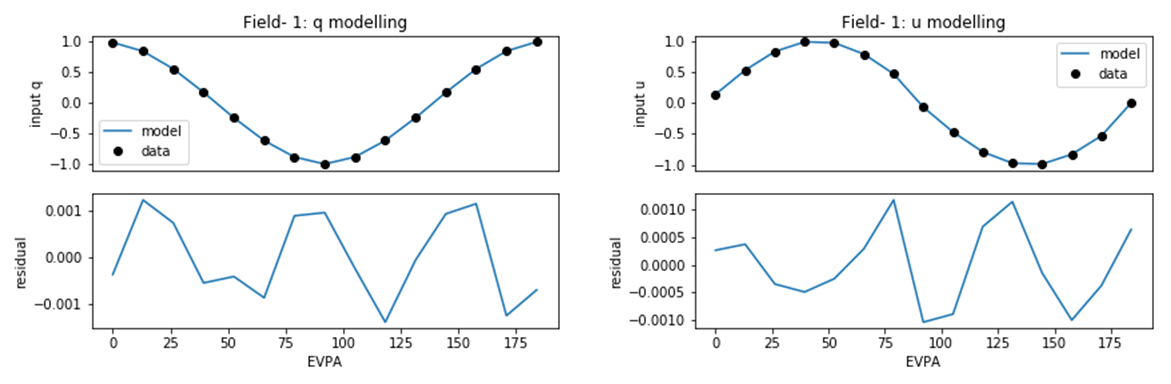}
\caption{Fitting of $q_i$ (left) and $u_i$ (right) as function of $\theta$ (EVPA) for fully linearly polarized input source for grid point 1, as per Equations~\ref{calibration_eqn_qi} and \ref{calibration_eqn_ui}.}%$ of the source.}%# As can be seen from the color-bars, in these plots, q and u values range from $\pm0.1$ as this is the expected bound of stellar polarization that WALOPs will observe. If needed, this plot can be extended to cover q/u range $\pm1.0$.}
\label{model_qu_fitting}
\end{figure}

From the measured intensities at the four detectors for the 15 input states of fully polarized light, $q_{m}$ and $u_{m}$ is found, as given by Equations~\ref{q_definition_cal} and \ref{u_definition_cal}. We use least squares method based curve fitting tools in Python to fit the Equations~\ref{calibration_eqn_qi} and \ref{calibration_eqn_ui} and obtain the coefficients. Fig~\ref{model_qu_fitting} shows the fit for grid point 1. To test the accuracy of the obtained calibration model, we need to verify that this model can predict the real Stokes parameters for partially polarized and unpolarized sources. A crucial requisite is the \textit{transmission model} of the instrument (described below in Section~\ref{transmission_model_appendix}), which accurately predicts the transmission at the detectors for any input Stokes vector to the telescope and instrument system. Each grid point is given as input 100 known $q$ and $u$ states of partially polarized light, with $p$ up to 5~\% and random $\theta$. Using the instrument \textit{transmission model}, we obtain the $N_{0}$, $N_{1}$, $N_{2}$ and $N_{3}$ and subsequently the $q_m$ and $u_m$ for these $q$ and $u$ values. Thereafter, we employ the calibration model to predict the input Stokes parameters, $q_i$ and $u_i$, from $q_m$ and $u_m$. Although we have used 100 test sources, in real observations, 5-10 sources (these can be standard polarized stars on sky) with near uniform spread in $q-u$ plane should be sufficient for testing the accuracy of the model.

\begin{figure}
\includegraphics[width = 1\linewidth]{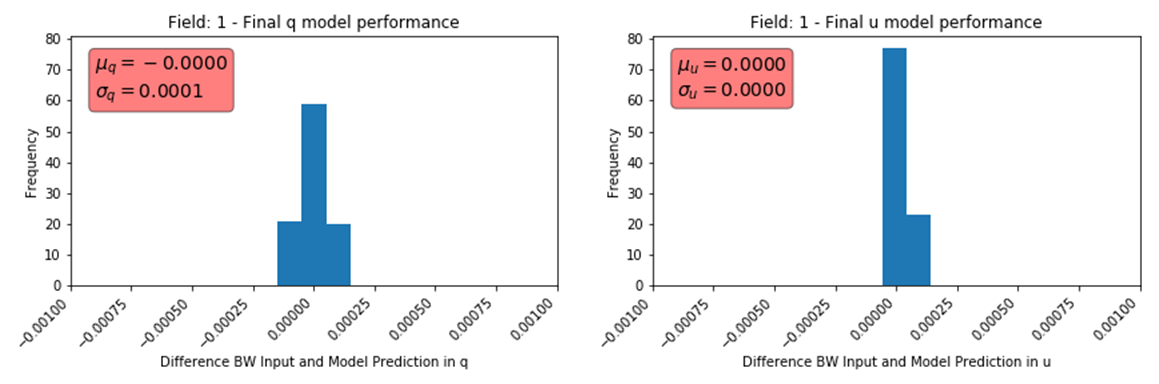}
\caption{Calibration model performance: Histogram plot of the difference between input and calibration model predicted $q$ (left) and $u$ (right) for 100 randomly partially polarized sources for grid point 1. Real values of both the Stokes parameters $q$ and $u$ can be recovered from the instrument measured values with accuracies better than 0.01~\% after the calibration model has been applied.}
\label{quhist_corrected}
\end{figure}

%To test the accuracy of the calibration model, each grid point is given as input 100\footnote{Here we use 100 test sources, while in real observations, 5-10 sources with near uniform spread in $q-u$ plane should be sufficient for testing the accuracy of the model.} known and random $q_i$ and $u_i$ states of partially polarized light\footnote{These will be standard polarized stars on sky.}, with $p$ up to 5~\% and random $\theta$. Using the instrument \textit{transmission model}, we obtain the $q_m$ and $u_m$ for these $q_i$ and $u_i$ values. Thereafter, we employ the calibration model to predict the input Stokes parameters from $q_m$ and $u_m$.% are used by the  to and check if we can retrieve these $q_i$ and $u_i$ values by using the model using the method describe above. 

For each grid point, we find a constant difference between the input and calibration model predicted Stokes parameters- between $q$ and $q_i$, and $u$ and $u_i$. That is, the calibration model created by using only fully polarized light is able to accurately predict all the coefficients in Equations~\ref{calibration_eqn_qi} and \ref{calibration_eqn_ui}, except for the $a_i$ parameters that are associated with unpolarized light. This is most likely due to the over-fitting of the Equations~\ref{calibration_eqn_qi} and \ref{calibration_eqn_ui} as we are using only fully polarized light to find the coefficients. This is corrected by calculating the mean difference between $q$ and $q_i$, and $u$ and $u_i$ of the 100 stars (to be referred to as $\Delta_q$ and $\Delta_u$ from hereon) and then applying this correction to their predicted $q_i$ and $u_i$ values. $\Delta_q$ and $\Delta_u$ are field dependent and can be up to 1~\% at some field points. The model performance for $q$ and $u$ is then estimated for these 100 sources after correction. The performance of the calibration model at each grid point can be characterized by the following four parameters- ${\mu}_q$, ${\mu}_u$, ${\sigma}_q$ and ${\sigma}_u$. These are the mean offset (difference with respect to real values) and spread in the recovery of the $q$ and $u$ parameters of the sources. The parameters ${\sigma}_q$ and ${\sigma}_u$ can be considered as the accuracy of the model. Fig~\ref{quhist_corrected} shows the histogram plot of the difference between the input and calibration model predicted Stokes parameters for grid point 1. The mean offset and spread associated with the recovery of the corresponding Stokes parameter are marked in their corresponding subplots. For the grid point 1, using the calibration model, the real Stokes parameters $q$ and $u$ can be recovered with accuracy of 0.01~\%. %\footnote{This can also be corrected by observing unpolarized standard sources.} So, for any grid point, irrespective of the value $q$ and $u$ given to the instrument, there is a constant difference between $q$ and $q_i$, and $u$ and $u_i$.

\subsection{Instrument Transmission Model} \label{transmission_model_appendix}
The \textit{polarization transmission} tool in Zemax software can only predict transmissions for fully polarized and unpolarized light. We created a \textit{transmission model} for all the four beams to accurately predict the fractional transmission value, $T_{k}$ (k = 1, 2, 3, 4) for each camera for any state of input polarized light to the system. Using the transmission data obtained from the Zemax software for the WALOP-South optical model for various states of polarized and unpolarized light, similar to the kind used for creating the calibration model, we create the transmission model. $T_{k}$ for any of the four beams can be written as a second order polynomial function of the intrinsic $q$ and $u$ value of the source, as given by Equation~\ref{t_eqn}. %depeFollowing are the steps in creating the transmission model:

\begin{equation}\label{t_eqn}
   T_{k} = s_{k1} + s_{k2}q + s_{k3}u + s_{k4}q^{2} + s_{k5}u^{2}
\end{equation}
 
\par
\begin{figure}
\centering
\includegraphics[width = 1\linewidth]{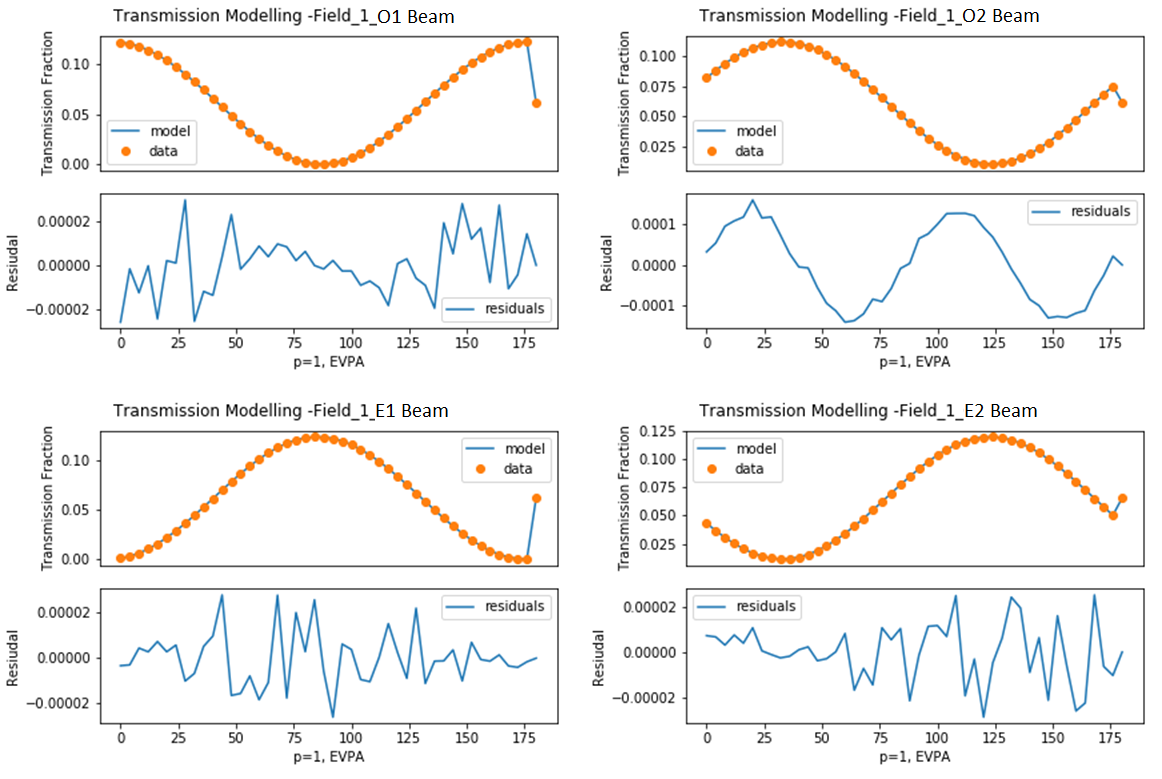}
\caption{Modelling of transmission for the four beams as function of various EVPA for fully polarized light as well as unpolarized light (extreme right point in the plots). The bottom subplots for each beam show the difference between the model and Zemax transmissions for the four beams.}%polarization. From this, the parameters $a_{i}$, $b_{i}$ and $c_{i}$ will be found. $a_{i}$ is the predicted transmission for the corresponding channel for unpolarized light beam. The last data point compares} transmission model's predicted unpolarized light transmission to that obtained from Zemax
\label{modelling1to4}
\end{figure}

 
While the transmission model has been developed for all the 144 grid points (Fig~\ref{grid_points_pol_input}), we present here the results for the grid point 1, which is representative of results from all other grid points. Fig~\ref{modelling1to4} shows the results (fits and residuals) of the transmission functions for the four beams. As can be seen, the fit is very good for both the polarized and unpolarized light beams with a residual of 0.02\% of the measured intensity.

\subsection{Calibration Model Results}
%\subsection{One-Shot Four Channel Polarimetry Zemax Model}

\begin{figure}
 \centering
 \includegraphics[width=1\linewidth]{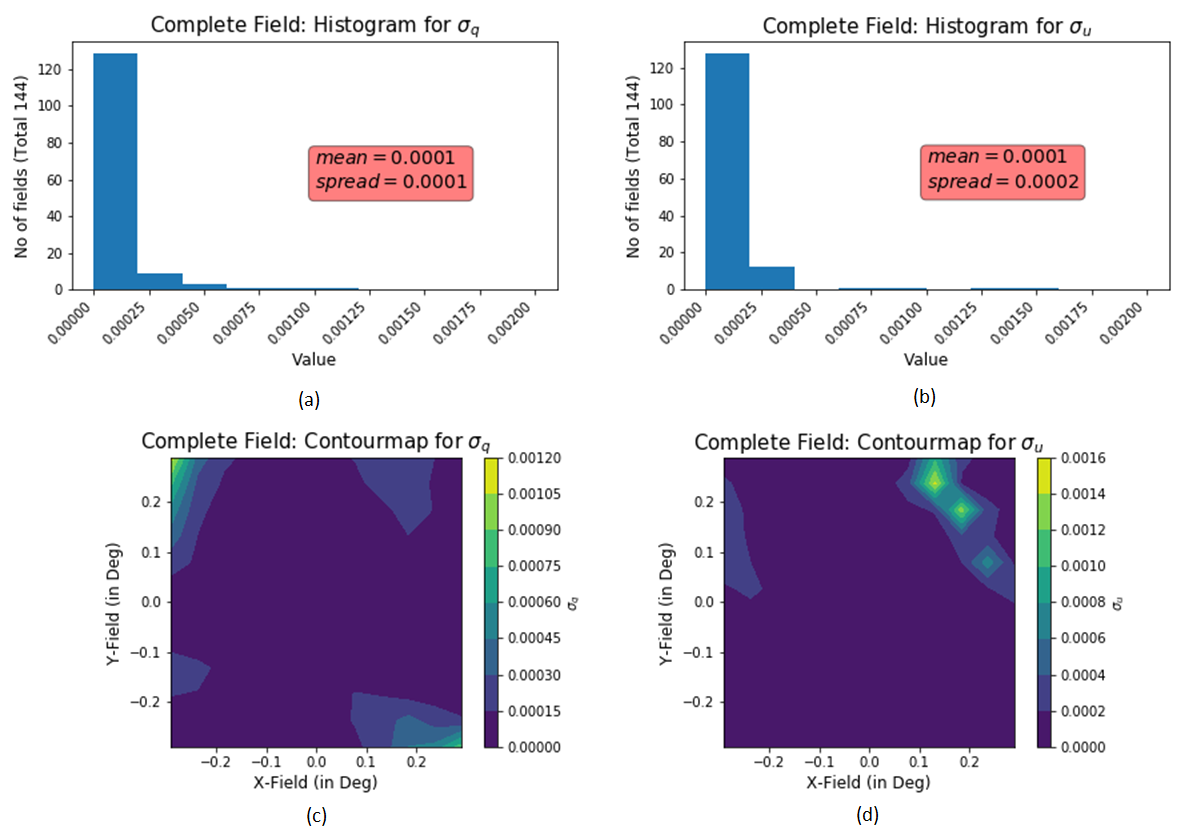}
  \caption{Performance of the calibration model for WALOP-South FoV: histogram (top) and contour plot (bottom). Left and right halves correspond to the results for the calibration model's accuracy for the $q$ and $u$ parameters, respectively.}
  \label{m12recovery}
  \end{figure}

\begin{figure}
\centering
\includegraphics[width=1\linewidth]{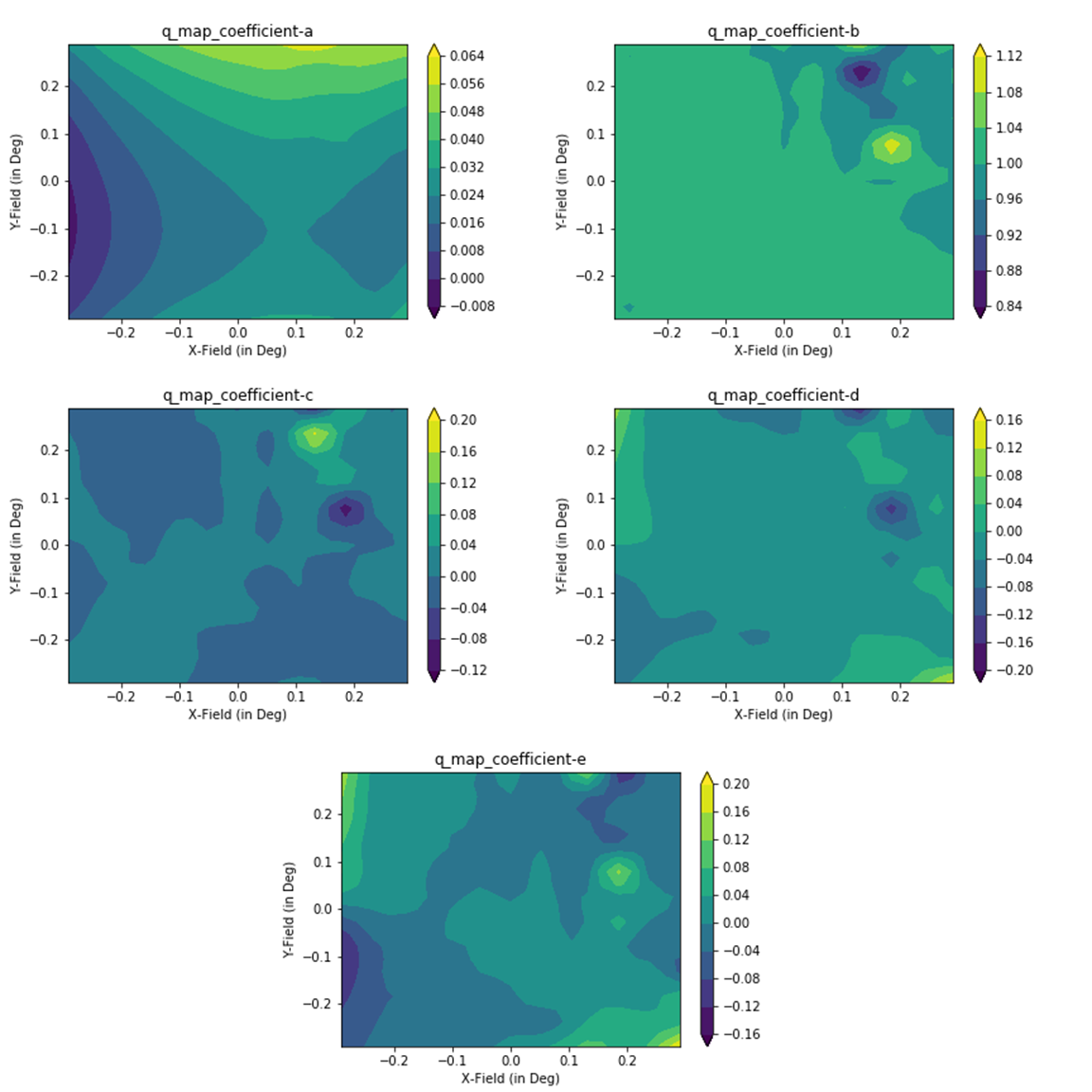}
\caption{Coefficients of the mapping function for $q$ parameter of the calibration model.}
\label{qm_coefficients}
\end{figure}

\begin{figure}
\centering
\includegraphics[width=1\linewidth]{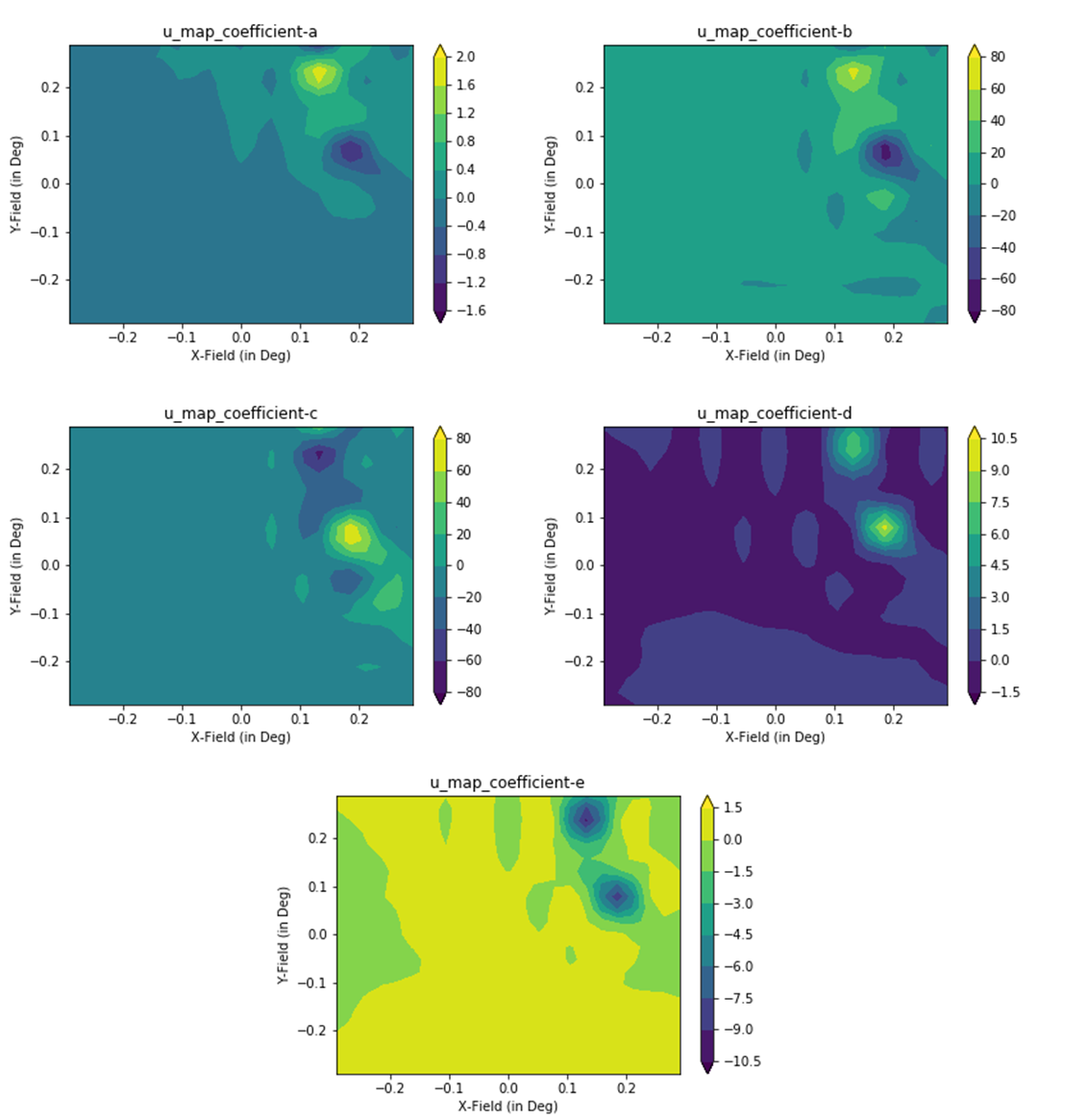}
\caption{Coefficients of the mapping function for $u$ parameter of the calibration model.}
\label{um_coefficients}
\end{figure}

The calibration model's results for grid point 1 shown in Fig~\ref{quhist_corrected} is representative of the model's performance across the FoV. We can recover the real Stokes parameters from the measured Stokes vector with high accuracy. The histogram and contour plot for $q_{\sigma}$ and $u_{\sigma}$ quantities for all the 144 grid points over the entire FoV are shown in Fig~\ref{m12recovery}. As can be seen from the results, the accuracy of the model for both $q$ and $u$ is better than 0.1~\% across the FoV, barring the strip where the cross-talk in $u_m$ from $q$ tends close to 1 and $u_{m}$ does not contain any contribution from $u$. This constitutes around 15~\% of the FoV area. Even in these regions, the accuracy of recovery is about 0.2\% at worst, which may compare reasonably with the expected photon noise induced uncertainties of the really faint stars in the field. Figs~\ref{qm_coefficients} and \ref{um_coefficients} show the values of the various coefficients of the mapping functions (Equations~\ref{calibration_eqn_qi} and \ref{calibration_eqn_ui}) across the FoV. As will be noticed, there are steep changes in the regions of high cross-talk.% Discussion on the results of the calibration model. Some regions are very well calibrated. How do the coefficients look across the FoV. There are some regions of high errors. These are the same regions with enormously high cross-talk.  

\subsubsection{Discussion}
The results from the previous section demonstrate that we can carry out high-accuracy linear polarimetry with the WALOP instruments across the FoV using the proposed calibration method. As Equations~\ref{calibration_eqn_qi} and \ref{calibration_eqn_ui} are second-order polynomial functions of ${q}_{m}$ and ${u}_{m}$, there is no degeneracy in the mapping of values from the measured ${q}_{m}$-${u}_{m}$ plane to the $q_i$-$u_i$ plane. In general, there is a larger error in the recovery of the $u$ parameter than that for $q$ for the entire field. This is expected due to the field-dependent cross-talk introduced in the $u$ measurement by the HWP in the WPA. During the commissioning as well as the operational lifetime of the instruments, the calibration model will be created and updated at frequent intervals.

For creating the calibration model, a source/star of any polarization can be used, as the CLP in the optical path ensures the input Stokes vector is only dependent on the CLP orientation angle. On the other hand, testing the accuracy of the calibration model requires multiple sources of known polarization, referred to as standard stars. As may be indicative from Figs~ \ref{qm_coefficients} and \ref{um_coefficients}, from  further analysis, it is evident that a calibration model based on 144 grid points can not be used to interpolate the model parameters across the entire FoV of WALOP-South. There is a steep change in the value of the coefficients in regions associated with high cross-talk, leading to inaccurate interpolated values for the coefficients in Equations~\ref{calibration_eqn_qi} and \ref{calibration_eqn_ui}. Therefore a finer spatial sampling of the FoV is required by increasing the number of grid points. This can be easily implemented using stellar fields with a high spatial density such as star clusters that fill the entire FoV, e.g., as implemented by Clemens et al. for calibrating the Mimir instrument for the Galactic Plane Infrared Polarization Survey\cite{clemens_GPIPS_Calibration}. However, making observations of 5-10 individual standard stars for each grid point would require a substantial amount of telescope time.

 \par The calibration model described above assumes a flat spectrum in the SDSS-r filter for the calibrating source as well as the science targets. In the sky, stars can have a significant difference in their spectral shape within the SDSS-r filter, owing to their different effective temperature. The instrument performance for different source spectra and mitigation strategies are described in Appendix~\ref{specpol_appendix}.%This and other aspects of on-sky implementation of the calibration model is discussed in the next section.Further, the model will be   Using the calibration model, for 12 magnitude stars, we can recover the $q$ and $u$ values with $0.01\pm0.04\%$ and $0.00\pm0.04\%$ accuracy respectively .% For 20 magnitude stars, the calibration model performance degrades due to large photon noise induced fitting errors.

    %\item By using the above calibration model, we can find the dependence of $q$ and $u$ on the measured ${q}_{m}$ and ${u}_{m}$, which turns out to be second order polynomial functions, as shown in Equation~\ref{eqn_q} and \ref{eqn_u}.

%The calibration model accounts for sky changes and photon noise from star and sky during the measurements.
%\textbf{The accuracy of the calibration model can be validated any time on sky by measuring the q and u of a standard polarized star and comparing it to that predicted by the model.} 
    %\item 4 out of 144 grid points have a large q-u recovery offset. We are trying to understand the reason behind it.
%The calibration scheme assumes we will be using unpolarized/polarized star clusters. So we have to identify suitable star clusters/fields on sky.

\iffalse
\begin{enumerate}
    \item \textit{Step 1}: Using observations through the CLP, parameters of the Equations~\ref{calibration_eqn_qi} and ~\ref{calibration_eqn_ui} are estimated. This step provides accurate estimate of all the cross-talk and efficiency coefficients, but not the zero-offset terms ($a_i$).  
    
    \item \textit{Step 2}: Observe 5-10 sources of known partial polarization or zero polarization, and accurately obtain the zero-offset terms ($a_i$) using the difference between the model predicted and real Stokes vector of the sources.
\end{enumerate}
\fi

\subsection{On Sky Implementation of the Calibration Scheme}
\par
\begin{figure}
\centering
\includegraphics[scale = 0.6]{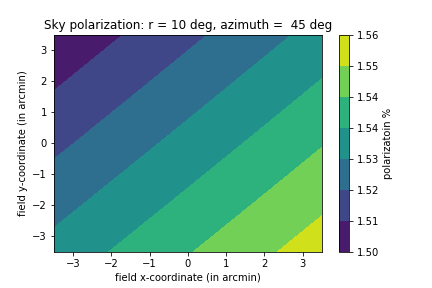}
\caption{Simulated polarization values for a 7 arcminutes square box at 10 degrees away from Moon and at azimuth angle of 45 degrees with respect to the Moon in a spherical coordinate system centered on the observer/telescope.}
\label{sky_pol_plot}
\end{figure}

 The main challenge in calibrating the WALOP polarimeters comes from their unprecedentedly large FoV. To calibrate extended FoV instruments, most often a raster-scanning method is used, in which the calibration model is created for a grid of points and interpolated for the entire FoV. For example, this was the approach that was used in RoboPol \cite{robopol}. Such an approach can work for a polarimeter requiring few grid points- i.e., if (i) it has a relatively small FoV of only a few arcminutes, and (ii) its polarimetric behaviour across the FoV is relatively simple that interpolation will suffice. For the WALOP polarimeters, due to their very large FoV and the relatively low noise floor requirements, the  raster-scanning method becomes unfeasible. Hence, we need a method that can calibrate the entire FoV with spatial continuity and minimal observation time. Gonz\'alez-Gait\'an et. al. \cite{moonlight_calibration_VLT} had used the bright sky adjacent to the full-Moon as an extended source for calibrating the FORS2 polarimeter on the 8~m Very Large Telescope. Light entering the atmosphere from the Moon on a full-Moon night is unpolarized and subsequent polarization is introduced owing to the scattering angle between the observer/telescope and Moon's position. The amount and angle of polarization can be modelled by using single Rayleigh scattering equations \cite{Harrington_solar_Calibration, Strutt_moonlight_polarization} as given by Equation~\ref{moonlight_p_eqn}. The polarization value depends on the angular distance of the region from the Moon ($\gamma$), and the polarization angle depends on the angle to the perpendicular scattering plane. $\delta$ is an empirical parameter whose value depends on the sky conditions, and for clear cloudless nights, it is found to be around 0.8\cite{moonlight_calibration_VLT}. Equation~\ref{moonlight_p_eqn} predicts that within an area of few arcminutes ($10\times10$~arcminutes) and sky positions of up to 15-20 degrees away from the Moon, the polarization fraction will remain constant to a level of few hundredths of a per cent, and has been verified through polarimetric observations with RoboPol (Maharana et al., in preparation). While second order effects lead to deviation of the measured polarization value from that predicted by theory, the polarization is expected to remain constant in that patch. Fig~\ref{sky_pol_plot} shows the expected linear polarization plot for a sky patch of size 7 arcminutes at 10 degrees away from the Moon. Gonz\'alez-Gait\'an et al. used the full-Moon sky to calibrate the FORS2 polarimeter, which has a FoV of 7 arcminutes, to an accuracy better than 0.05~\%. We plan to use the full-Moon sky for carrying out both steps of the calibration model development.  % for calibrating polarimeters with large FoV due to time considerations. In addition, for WALOP, the polarimetric behaviour changes rapidly in some part of the FoV.  %are presented. %We will test the calibration model on sky during the commissioning of WALOP-South instrument and details of the model and results obtained will be published as a separate paper. We have also tested and verified the calibration method in a table top polarimeter in the lab.The theoretical and experimental results are consistent with each other.

\begin{equation}\label{moonlight_p_eqn}
p = \delta\frac{sin^{2}\gamma}{1 + cos^{2}\gamma}
\end{equation}
\iffalse
\begin{equation}\label{moonlight_theta_eqn}
\theta = cos^{-1}\frac{cos\gamma - cos\phi cos\psi}{sin\phi sin\psi}
\end{equation}
\fi

\begin{figure}
    \centering
    \includegraphics[scale = 0.35]{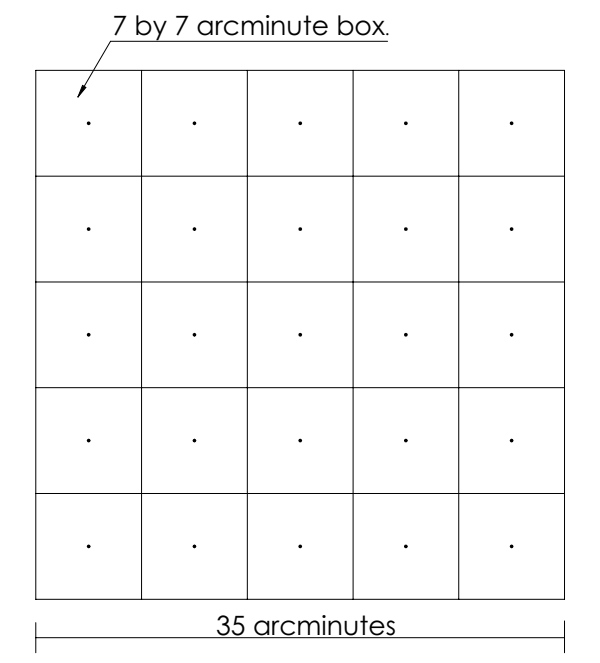}
    \caption{On sky calibration scheme for WALOP-South instrument. The entire FoV is divided into 25 boxes of size $7\times7$~arcminutes each. Each of these boxes will be calibrated as separate units.}
    \label{calibration_binning}
\end{figure}

The on-sky calibration of the WALOP instruments will be carried out in the following steps to obtain a spatially continuous calibration model. Firstly, the preliminary calibration model will be created for the entire FoV by observing bright wide-field/extended sources such as the star clusters, twilight or full-Moon sky through the CLP. The only requirement is that the polarization of the source should not change within the exposure time as that will lead to uncertainty in the polarization of the beam passing through the CLP. Using standard polarized stars on sky, determination and correction for $\Delta_q$ and $\Delta_u$, and subsequent estimation of calibration accuracy will be done at the centers of the 5x5 grid into which the entire FoV will be divided, as shown in Fig~\ref{calibration_binning}. This will provide us accurate calibration model for these points.  Subsequently, observation of the full-Moon sky patches at different (5-10) combinations of separation and azimuth angles with respect to the Moon will be used as standard polarized patches to carry out zero-offset correction as well as establish the calibration accuracy for the entire FoV. The uncertainty in the value of absolute polarization in each box is overcome by using the already developed calibration model at the central point of each box.

%Following the above steps, we can carry out complete calibration of the entire FoV of the WALOP instruments. 

\par From the calibration modelling results, we find a higher error ($> 0.1~\%$) in the calibration model performance for the $u$ parameter in a narrow patch where the cross-talk level is high. To compensate for this, between exposures, the calibration HWP inside the instrument (refer to Paper I) can be alternatively placed at position angles of $0^{\circ}$ and $45^{\circ}$. This will effectively interchange the $q$ and $u$ channels of the instrument. Consequently, by averaging the measurements from the two calibration HWP orientations, in regions of high cross-talk, the modelling errors can be reduced. %Additionally, the planning software for the \textsc{pasiphae} survey will use the polarmetric accuracy map of WALOPs across their FoV to optimize the survey footprint such that minimal number of stars on sky fall on these patches.

\section{Lab Set Up Tests}\label{calibration_model_lab_test}

\begin{figure}[htbp!]
\includegraphics[width=1\linewidth]{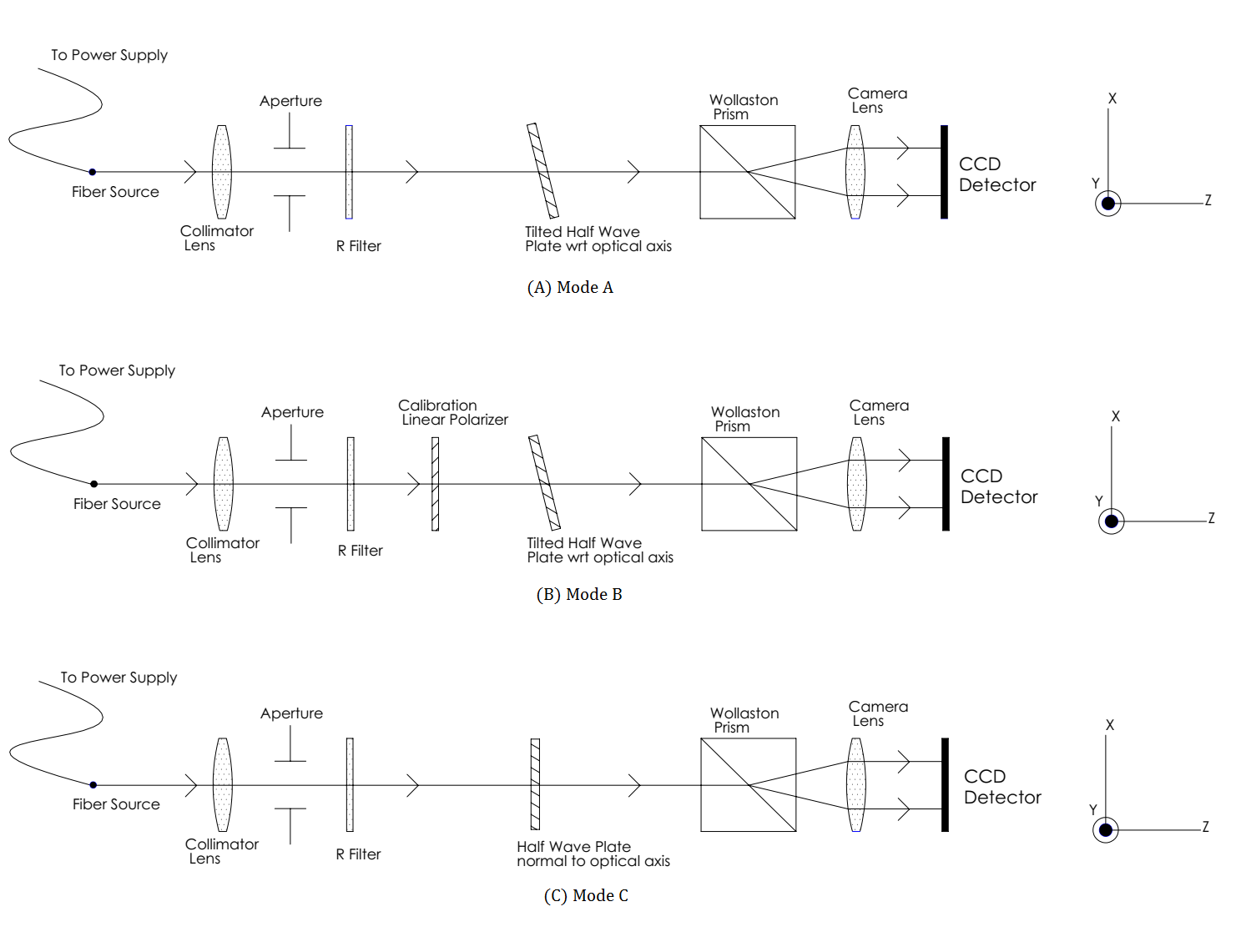}
\caption{Schematic of the test-bed polarimeter setup used for testing the performance of the WALOP calibration model. The polarimeter works in three modes- (a) main polarimeter set up with high level of cross-talk due to tilted HWP, (b) inserted calibration linear polarizer in the polarimeter to carry out calibration observations and (c) set-up with non-tilted HWP to measure the real polarization of sources used for testing the model accuracy. }
\label{lab_setup_overall}
\end{figure}

\begin{figure}
    \centering
    \includegraphics[width=1\linewidth]{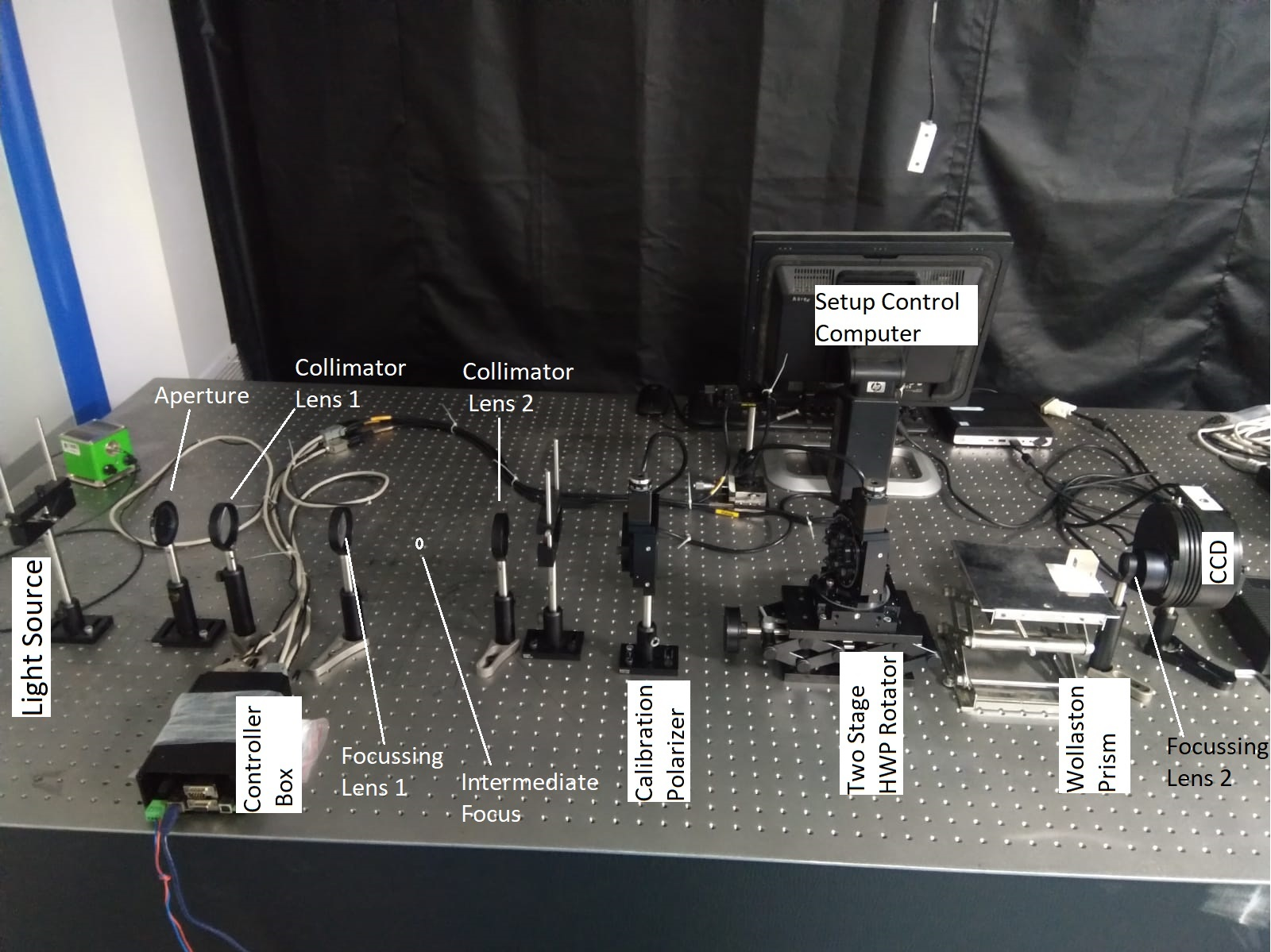}
    \caption{A table top polarimeter set-up in the lab to test WALOP Calibration Model.}
    \label{Set_up_pic}
\end{figure}

To test the calibration model on real WALOP-like polarimeter systems, a table-top test-bed polarimeter was set up in the lab in IUCAA. The schematic of the set-up along with its three modes of operation is shown in Fig~\ref{lab_setup_overall} and the actual lab setup is shown in Fig~\ref{Set_up_pic}. It is a dual-channel polarimeter consisting of a rotating HWP and a WP as the analyzer system. An LED-fed fiber source was used to simulate star like point source for the experiment. The HWP is placed with its normal tilted with respect to the optic axis of the system so that the  rays are incident on the HWP at oblique angles so as to create cross-talks similar to what is expected in WALOPs. For measuring $q$, the HWP's fast axis is aligned with the x-axis of the Instrument Coordinate System (ICS) and for $u$, it is oriented at $22.5^{\circ}$. Combining these two measurements, we are able to simulate both the left and right halves of the WPA of the WALOP instruments, which is where all the polarimetric cross-talk originates. Various levels of cross-talk between 16~\% to 84~\% were simulated, as noted under the column \textit{Tilt Angle} in Table~\ref{Calibration_lab_Results}. The percentile area corresponding to each cross-talk level in the WALOP-South instrument is noted in the same Table. The operation of the rotation stages and the camera  were fully automated through an instrument control software written in Python and run through the Setup Control Computer (shown in Fig~\ref{Set_up_pic}). An SBIG-ST9 CCD camera was used as the detector. The photometric and subsequent polarimetric analysis of the data was carried out through a data reduction pipeline written in Python, with careful emphasis on error estimation and propagation in each step. For photometry, Photutils software\cite{photutils} was used for implementing aperture photometry. %The calibration of this polarimeter was done in the same way as proposed in the calibration model for WALOPs, through its three modes of operation.% Operation of the set up is shown in Figure~\ref{lab_setup_overall}}

\iffalse
\begin{subfigure}{0.99\textwidth}
\centering
\frame{\includegraphics[scale = 0.4]{Lab_setup_measured_polarization.png}}
\caption{ High cross-talk polarimeter system to create polarimetric cross-talk expected in WALOPs. The tilt angle of the HWP is controlled through a motorized rotary stage to introduce various cross-talk levels. Another rotation stage controls the orientation of the HWP fast-axis (on the X-Y plane).}
\label{lab_setup}
\end{subfigure}
\centering
\begin{subfigure}{0.99\textwidth}
\centering
\frame{\includegraphics[scale = 0.4]{Lab_setup_calibration.png}}
\caption{Calibration linear polarizer, mounted on a motorized rotation stage, is inserted in the polarimeter to carry out calibration measurements. }
\label{lab_setup_calibration}
\end{subfigure}
\centering
\begin{subfigure}{0.99\textwidth}
\centering
\frame{\includegraphics[scale = 0.4]{Lab_setup_intrinsic_polarization.png}}
\caption{To measure real linear polarization values of partially polarized sources used in the calibration model, HWP is oriented normal to optical axis and standard two-channel polarimetry is carried out by rotating it around the optical axis. }
\label{lab_setup_intrinsic}
\end{subfigure}
\fi

\par
\begin{figure}
    \centering
    \includegraphics[scale = 0.65]{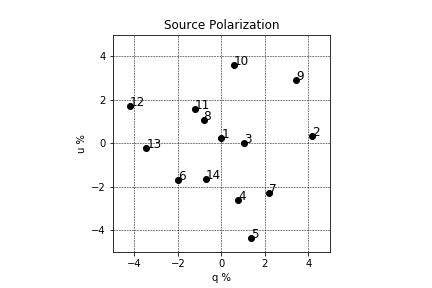}
    \caption{Polarization of sources used for testing the accuracy of the calibration model in the lab.}
    \label{calibration_test_sources}
\end{figure}

 The calibration of this polarimeter was done in the same way as proposed in the calibration model for WALOPs, through its three modes of operation. At the beginning of this system, a calibration linear polarizer mounted on a motorized rotation stage is placed in the optical path during calibrations, as shown in Fig~\ref{lab_setup_overall}~(b) and removed thereafter. The polarimeters at different HWP tilts were treated as independent systems and thus the calibration was done independently for them. Calibration \textit{step 1} is done by inserting the calibration linear polarizer in the optical path and feeding 15 states of fully polarized light as input for which the corresponding $q_m$ and $u_m$ were measured. Using this, the calibration equations were fitted and coefficients were obtained. Once the \textit{step 1} of the model is executed, we need to carry out \textit{step 2} to test the accuracy of the model. For this, multiple partially polarized beams were created in the set up by placing an old polarizer film sheet at an intermediate focus position before the aperture of the instrument (Figs~\ref{Set_up_pic} and \ref{lab_setup_overall}~(a)). The polarizer's coating had worn off and effectively became a partial polarizer sheet. Using this sheet, polarization values between 0 and 5~\% were generated, depending on the spatial position of incident light on the sheet. We checked the stability of the strip by making repeated polarization measurements over time. Fig~\ref{calibration_test_sources} shows the $q-u$ values generated using this method. To measure the intrinsic polarization of these beams created through this method, we corrected for the tilt of the HWP by making it normal to the optical axis through the rotary stage on which it is mounted, as shown in Fig~\ref{lab_setup_overall}~(c). Using conventional two-channel polarimetry with this polarimeter system, the intrinsic polarization of the sources were found ($q$, $u$, and $p$), to have accuracies better than 0.03~\%. The $q_m$ and $u_m$ measured with the tilted HWP systems were then used in the calibration model to predict the $q_i$ and $u_i$ (and $p_i$ using these) for each of the tilted polarimeters. As found in the case of the theoretical calibration model for the WALOP-South instrument, just by using fully polarized light in \textit{step 1}, all the coefficients could be accurately estimated barring the zero-offset terms ( $a_i$ terms in Equations~\ref{calibration_eqn_qi} and \ref{calibration_eqn_ui}). The mean offset in $q$ and $u$ between the calibration model predicted and real Stokes parameters ($\Delta_q$ and $\Delta_u$ ) were corrected and the calibration accuracy was estimated.

\begin{figure}
    \centering
    \includegraphics[scale = 0.46]{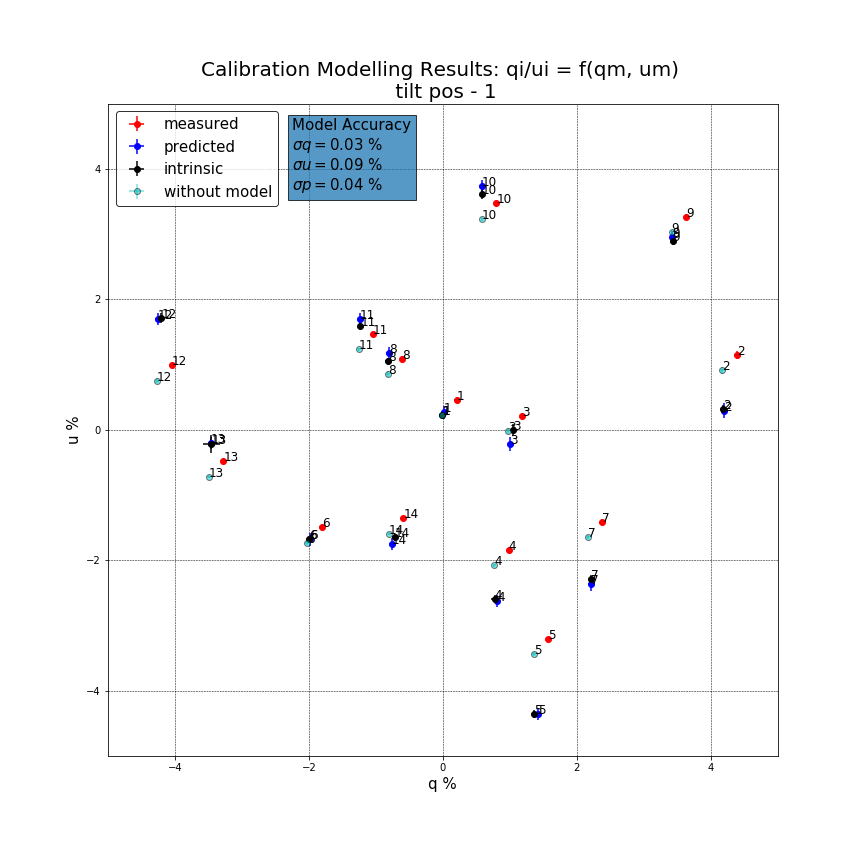}
    \caption{Performance of the calibration model in calibrating WALOP test-bed polarimeter with cross-talk value of 16~\%, corresponding to a percentile area of 48.5~\% over WALOP-South FoV.}
    \label{calibration_results_iqu}
\end{figure}

\begin{figure}
\centering
\includegraphics[width=1\linewidth]{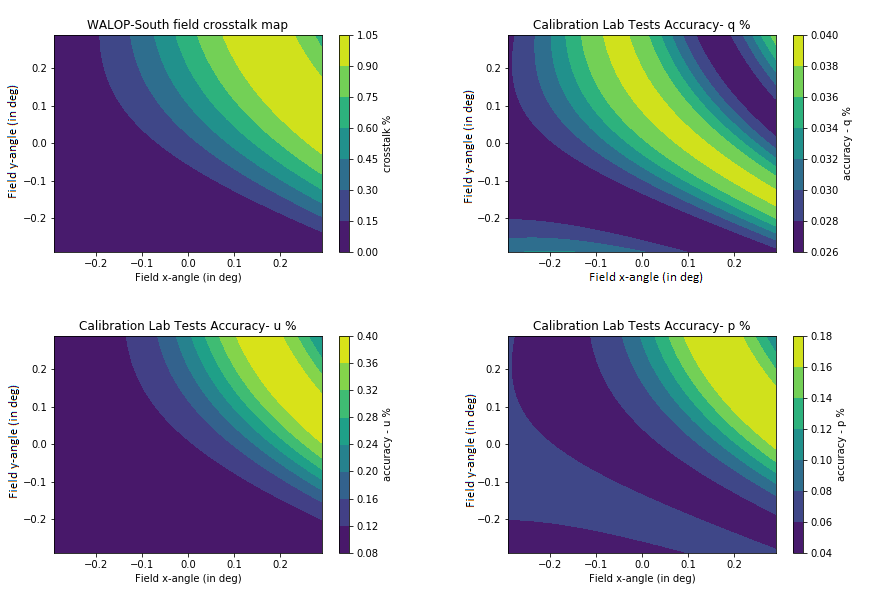}
\caption{Predicted accuracy of calibration model for $q$, $u$ and $p$ over WALOP-South FoV based on lab testing results of the calibration model.}
\label{lab_results_WALOP_FoV}
\end{figure}

%Sources of polarization values between 0 and 5~\% spread uniformly over the $q-u$ plane, as shown in Fig~\ref{calibration_test_sources}, were created in the setup.
Fig~\ref{calibration_results_iqu} plots the results for one of the HWP tilts leading to a cross-talk value of 16~\%, corresponding to around 50~\% area percentile of WALOP-South FoV. While the black and blue points correspond to the real and model predicted $q-u$ values, respectively, the red points are the instrument-measured Stokes parameters ($q_m$ and $u_m$). The gray point is the estimated $q-u$ prediction after correction for only zero-offset, which is the standard practice in most polarimeters since their cross-talk is nearly zero. As can be seen, without polarimetric cross-talk correction, the predicted polarization is very inaccurate. Once the measured $q-u$ values are corrected with the calibration model, the model predicted and intrinsic polarization values match to better than 0.04~\%. For estimating accuracy in $p$, $p_i$ derived from the calibration model predicted $q_i$ and $u_i$ is compared to the intrinsic $p$ of the source. As error in $p$ is a weighted average of errors in $q$ and $u$ (~$\sigma_{p} = \sqrt{\frac{q^{2}{\sigma_{q}^{2}}+ u^{2}{\sigma_{u}^{2}}}{q^{2} + u^{2}}}$), we get an overall polarimetric accuracy in $p$ lying between accuracy of $q$ and $u$. Likewise, there is an asymmetric contribution from errors in $q$ and $u$ to EVPA ($\theta$) measurement errors. Other cross-talk values (created using different tilt angles of the HWP) and the corresponding performance of the calibration model are shown in Table~\ref{Calibration_lab_Results}. Fig~\ref{lab_results_WALOP_FoV} is the expected polarimetric accuracy across the WALOP-South FoV based on the instrument's cross-talk map. In summary, we are able to calibrate more than 75~\% of the WALOP-South instrument's FoV with better than 0.1~\% accuracy in $p$ and the remaining area to better than 0.2~\%.

%\subsubsection{Results}

\begin{table}[]
\begin{center}

\begin{tabular}{ccccccc}
\hline
\multicolumn{1}{|c|}{\begin{tabular}[c]{@{}c@{}}Position\\ No\\ \end{tabular}} &
\multicolumn{1}{c|}{\begin{tabular}[c]{@{}c@{}}Tilt\\ Angle\\ \end{tabular}} &
\multicolumn{1}{c|}{\begin{tabular}[c]{@{}c@{}}Cross-talk\\ Level\\ \end{tabular}} & \multicolumn{1}{c|}{\begin{tabular}[c]{@{}c@{}}Calibration\\ Accuracy\\ $q$\end{tabular}} & \multicolumn{1}{c|}{\begin{tabular}[c]{@{}c@{}}Calibration\\ Accuracy\\ $u$\end{tabular}} & \multicolumn{1}{c|}{\begin{tabular}[c]{@{}c@{}}Calibration\\ Accuracy\\ $p$\end{tabular}} & \multicolumn{1}{c|}{\begin{tabular}[c]{@{}c@{}}Percentile Area \\ of WALOP \\ field of view\end{tabular}} \\ \hline
\multicolumn{1}{|c|}{1}                & \multicolumn{1}{c|}{12}         & \multicolumn{1}{c|}{16 \%}            & \multicolumn{1}{c|}{0.03 \%}                                                            & \multicolumn{1}{c|}{0.09 \%}                                                            & \multicolumn{1}{c|}{0.04 \%}                                                            & \multicolumn{1}{c|}{48.5 \%}                                                                              \\ \hline
\multicolumn{1}{|c|}{2}                & \multicolumn{1}{c|}{14}         & \multicolumn{1}{c|}{27 \%}          & \multicolumn{1}{c|}{0.04 \%}                                                            & \multicolumn{1}{c|}{0.14 \%}                                                            & \multicolumn{1}{c|}{0.08 \%}                                                            & \multicolumn{1}{c|}{58 \%}                                                                                \\ \hline
\multicolumn{1}{|c|}{3}                & \multicolumn{1}{c|}{16}         & \multicolumn{1}{c|}{43 \%}          & \multicolumn{1}{c|}{0.04 \%}                                                            & \multicolumn{1}{c|}{0.16 \%}                                                            & \multicolumn{1}{c|}{0.07 \%}                                                            & \multicolumn{1}{c|}{67 \%}                                                                                \\ \hline
\multicolumn{1}{|c|}{4}                & \multicolumn{1}{c|}{18}         & \multicolumn{1}{c|}{62 \%}            & \multicolumn{1}{c|}{0.04 \%}                                                            & \multicolumn{1}{c|}{0.17 \%}                                                            & \multicolumn{1}{c|}{0.07 \%}                                                            & \multicolumn{1}{c|}{75.6 \%}                                                                              \\ \hline
\multicolumn{1}{|c|}{5}                & \multicolumn{1}{c|}{20}         & \multicolumn{1}{c|}{81 \%}            & \multicolumn{1}{c|}{0.03\%}                                                             & \multicolumn{1}{c|}{0.29 \%}                                                            & \multicolumn{1}{c|}{0.13 \%}                                                            & \multicolumn{1}{c|}{83.9 \%}                                                                              \\ \hline
\multicolumn{1}{l}{}                   & \multicolumn{1}{l}{}            & \multicolumn{1}{l}{}                  & \multicolumn{1}{l}{}                                                                    & \multicolumn{1}{l}{}                                                                    &                                                                                   
\end{tabular}
\end{center}
\caption{Results from the lab calibration tests of the WALOP calibration model. For more than 75~\% of the WALOP-South instrument's field we obtain accuracies better than the target accuracy of 0.1~\% in $p$ (degree of polarization).}
\label{Calibration_lab_Results}
\end{table}

\iffalse
\begin{figure}
    \centering
    \frame{\includegraphics[scale = 0.25]{Overall_Results.png}}
    \caption{Calibration Results Overall}
    \label{calibration_lan_Results_plot}
\end{figure}
\fi

\section{Conclusions}\label{calibration_conclusion}

We have carried out complete polarimetric analysis of the WALOP-South instrument's optical system to estimate the polarimetric cross-talk in the measurements of the Stokes parameters. We find significant cross-talk in the measured Stokes parameters. Almost all the cross-talk originates from the HWPs in the WPA, owing to the large ($> 5^\circ$) oblique angles of incidence. The effect is very well reproduced using ray tracing principles applied on the HWP. To correct for this, we have developed a complete on-sky calibration method, enabling us to obtain $0.1~\%$ accuracy across most of the FoV. The implementation of the calibration model on sky is relatively straight forward. This work was done using polarization analysis tools available in Zemax software as well as utilizing ray tracing programs developed in Python. Before fabrication and assembly of an instrument, the expected instrumental polarization can be estimated using the polarization analysis features in optical design software like \href{https://www.zemax.com/}{Zemax}\textsuperscript{\textregistered} or \href{https://www.breault.com/software/asap-capabilities}{Advanced Systems Analysis Program (ASAP)}\textsuperscript{\textregistered}. The tools developed for the polarization modelling and calibration of WALOPs can be used for any other polarimeter to understand the various polarimetric systematic effects induced by the optics and prepare the required calibration plan. Additionally, we tested the calibration method on a test-bed WALOP like polarimeter in the lab and find that we can carry out high accuracy polarimetric calibrations at various cross-talk levels. The calibration method will be implemented during the on-sky commissioning of the instrument and subsequent results will be published as part of the instrument commissioning paper. %This work was done using polarization analysis tools available in Zemax software as well as utilizing ray tracing programs developed in Python.

\appendix

\section{Measured Stokes Parameters and Instrument Matrix}\label{instrument_matrix_appendix}
The incident intensity at the detector for any polarization channel of the instrument ($0^\circ$, $45^\circ$, $90^\circ$ and $135^\circ$ polarizations) can be written as:
    \begin{equation}
    I_{\theta} = a_{\theta} + b_{\theta}q + c_{\theta}u 
\end{equation}
where $a_{\theta}$, $b_{\theta}$ and $c_{\theta}$ correspond to the elements of the first row of the Mueller matrix for the optical path corresponding to that polarization angle/channel. The normalized difference between intensities corresponding to two orthogonal polarization angles/channels yields $q$ or $u$, as given by the following equation.
\begin{equation}
    r_{i} = \frac{I_{\theta1} - I_{\theta2}}{I_{\theta1} + I_{\theta2}}
    = \frac{(a_{\theta1} + b_{\theta1}q + c_{\theta1}u) - (a_{\theta2} + b_{\theta2}q + c_{\theta2}u)} {(a_{\theta1} + b_{\theta1}q + c_{\theta1}u) + (a_{\theta2} + b_{\theta2}q + c_{\theta2}u)} 
\end{equation}
The binomial expansion of $(1+x)^{-1}$ is given by $(1+x)^{-1} = 1 - x +x^2 -x^3 + x^4$. Using that, the generalized normalized difference can be written as a polynomial equation in $q$ and $u$.
\begin{equation}
 r_{i} = A_{i} + B_{i}q + C_{i}u + D_{i}q^2 + E_{i}u^2 + F_{i}qu + ...
\end{equation}
In general, for most simple polarimeters, the second order terms are zero, and the instrument measured Stokes parameters can be written as a set of linear equations, together forming a Muller matrix like \textit{Instrument Matrix}. The first row is inconsequential to polarimetric measurements and hence can be ignored.
\begin{gather}%\label{mueller_matrix_instrument_normalized}
s_{m} = 
\begin{bmatrix}
1 \\
q_{m} \\
u_{m} \\
v_{m} 
\end{bmatrix}
= 
m_{inst}\times s
= 
\begin{bmatrix}
m_{11} & m_{12} & m_{13} & m_{14}  \\
m_{21} & m_{22} & m_{23} & m_{24}  \\
m_{31} & m_{32} & m_{33} & m_{34}  \\
m_{41} & m_{42} & m_{43} & m_{44} 
\end{bmatrix}
\times  
\begin{bmatrix}
1 \\
q \\
u \\
v 
\end{bmatrix}
\\
=
\begin{bmatrix}
- & - & - & -  \\
1 \,\to\,q_{m} & q\,\to\,q_{m} & u\,\to\,q_{m} & v\,\to\,q_{m}  \\
1 \,\to\,u_{m} & q\,\to\,u_{m} & u\,\to\,u_{m} & v\,\to\,u_{m}  \\
1 \,\to\,v_{m} & q\,\to\,v_{m} & u\,\to\,v_{m} & v\,\to\,v_{m}
\end{bmatrix}
\times  
\begin{bmatrix}
1 \\
q \\
u \\
v 
\end{bmatrix}
\end{gather}

\section{Polarimetric Modelling of HWPs}\label{HWP_modelling_section}
The Mueller Matrix of a retarder plate with retardance $\delta$ oriented at an angle $\alpha$ with respect to the Instrument Coordinate System (ICS) is given by Equation~\ref{genera_retarder_matrix}.% (more details in Appendix~\ref{Mueller_matrix_polarimeter}). %can be found by using Equations~\ref{mmatrix_rotation_algebra} and \ref{retarder_matrix_nominal}, and is given by Equation~\ref{retarder_matrix_rotated}.

\begin{gather}
M_{\alpha, \delta} = M_{rot}(-\alpha)\times M_{\delta} \times M_{rot}(\alpha) \nonumber \\
 = 
\centering
\begin{bmatrix}
1 & 0 & 0 & 0  \\
0 & cos^{2}2\alpha + sin^{2}2\alpha cos\delta & cos2\alpha sin2\alpha(1- cos\delta) & -sin2\alpha sin\delta   \\
0 & cos2\alpha sin2\alpha(1- cos\delta) & cos^{2}2\alpha cos\delta + sin^{2}2\alpha & cos2\alpha sin\delta   \\
0 & sin2\alpha sin\delta & -cos2\alpha sin\delta & cos\delta
\end{bmatrix}
\label{genera_retarder_matrix}
\end{gather}

Thus the Mueller matrices for the HWPs used in the WPA, to be referred as HWP1 (oriented at $0^{\circ}$ with ICS) and HWP2 (oriented at $22.5^{\circ}$ with ICS) from hereon, are given by Equations~\ref{retarder_matrix_non_ideal_LHWP} and \ref{retarder_matrix_non_ideal_RHWP}.

\begin{gather}\label{retarder_matrix_non_ideal_LHWP}
M_{0, \delta} = 
\begin{bmatrix}
1 & 0 & 0 & 0  \\
0 & 1 & 0 & 0 \\
0 & 0 & cos\delta & sin\delta    \\
0 & 0  & -sin\delta  & cos\delta
\end{bmatrix}
\end{gather}
\begin{gather}\label{retarder_matrix_non_ideal_RHWP}
M_{22.5, \delta} = 
\begin{bmatrix}
1 & 0 & 0 & 0  \\
0 & \frac{1}{2}(1 + cos\delta) & \frac{1}{2}(1-cos\delta) & -\frac{1}{\sqrt{2}} sin\delta  \\
0 & \frac{1}{2}(1-cos\delta) &\frac{1}{2}(1 + cos\delta) & \frac{1}{\sqrt{2}} sin\delta    \\
0 & \frac{1}{\sqrt{2}} sin\delta  & -\frac{1}{\sqrt{2}} sin\delta  & cos\delta
\end{bmatrix}
\end{gather}

%The general Stokes vector of a beam after passing through a HWP is given by Equation~\ref{Svector_general}. Applying this to HWP1 and HWP2 yields the Stokes vector as given in Equation~\ref{Svector_nonideal_LHWP} and \ref{Svector_nonideal_RHWP}.

Hence the Stokes vector of a beam after passing through the two HWPs are as given in Equation~\ref{Svector_nonideal_LHWP} and \ref{Svector_nonideal_RHWP}.

\iffalse
\begin{gather}\label{Svector_general}
S_{\alpha, \delta} =
\frac{1}{2}
\begin{bmatrix}
1 & 0 & 0 & 0 \\
0 & M_{22} & M_{23} & M_{24}  \\
0 & M_{32} & M_{33} & M_{34}  \\
0 & M_{42} & M_{43} & M_{44} 
\end{bmatrix}
\times
\begin{bmatrix}
1  \\
q  \\
u  \\
v 
\end{bmatrix}
= 
\begin{bmatrix}
1   \\
M_{22}{q} + M_{23}{u} + M_{24}{v}  \\
M_{32}{q} + M_{33}{u} + M_{34}{v}    \\
M_{42}{q} + M_{43}{u} + M_{44}{v}
\end{bmatrix}
\end{gather}
\fi

\begin{gather}\label{Svector_nonideal_LHWP}
s_{0, \delta} =
\begin{bmatrix}
1  \\
q  \\
u cos\delta + v sin\delta    \\
-u sin\delta  + v cos\delta
\end{bmatrix}
\end{gather}

\begin{gather}\label{Svector_nonideal_RHWP}
s_{22.5, \delta} =
\begin{bmatrix}
1   \\
\frac{q}{2}(1 + cos\delta) +\frac{u}{2}(1-cos\delta)  -\frac{v}{\sqrt{2}} sin\delta  \\
\frac{q}{2}(1-cos\delta) + \frac{u}{2}(1 + cos\delta) + \frac{v}{\sqrt{2}} sin\delta    \\
\frac{q}{\sqrt{2}} sin\delta  -\frac{u}{\sqrt{2}} sin\delta  + v cos\delta
\end{bmatrix}
\end{gather}

The general formula for the measured Stokes parameter for a two-channel or four-channel polarimeter using a Wollaston Prism that splits $0^{\circ}$ and $90^{\circ}$ polarizations corresponds to the q-element of the Stokes vector as given by Equation~\ref{r_non_ideal}, from which the  $q_{m}$ and $u_{m}$ for can be found using Equations~\ref{q_non_ideal} and \ref{u_non_ideal}. As can be seen, the $q_{m}$ is unaffected by non-half wave retardance and depends only on $q$ of the source while $u_{m}$ can have a strong dependence on  all the intrinsic Stokes parameters $q$, $u$ and $v$ through the Mueller matrix elements. Only in the case of retardance of $\lambda/2$, $u_{m}$ equals $u$. As an extreme example, if the retardance is $\lambda/4$ instead of  $\lambda/2 $, $u_{m}$ equally depends on $q$, $u$ and $v$ as shown by Equation~\ref{u_non_ideal_quarter_wave}.

\begin{equation}\label{r_non_ideal}
    r_{m} = m_{22}{q} + m_{23}{u} + m_{24}{v}  
\end{equation}

\begin{equation}\label{q_non_ideal}
    q_{m} = r(0, \delta) = q
\end{equation}

\begin{equation}\label{u_non_ideal}
    u_{m} = r(22.5, \delta) = \frac{q}{2}(1 + cos\delta) +\frac{u}{2}(1-cos\delta)  -\frac{v}{\sqrt{2}} sin(\delta)
\end{equation}

\begin{equation}\label{u_non_ideal_quarter_wave}
    u^{'}_{m} = r(22.5, \lambda/4) = \frac{q}{2} +\frac{u}{2} -\frac{v}{\sqrt{2}}
\end{equation}

%The Mueller matrix of a retarder plate aligned along the instrument's x-axis is given by half-wave plate is given by Equation~\ref{mmat_ret}\cite{collet}:

Assuming circular polarization as zero, the formula for $u_{m}$ can be written as Equation~\ref{u_non_ideal2}. $m_{22}$ term captures the cross-talk from $q$ to $u_{m}$ and the $m_{23}$ term captures the fraction of $u$ converted into $u_{m}$. $u_{m}$ is dependent on the only on retardance apart from the input Stokes vectors. For a light ray, retardance depends on the incident angle (angle with the normal) and the azimuth angle with the HWP. For any point in the WALOP-South FoV, both these angles can be found using the Lagrange invariant equation (refer to Paper I) at the pupil and from thereon using Snell's law based ray tracing equations after the BK7 wedges and incident on the HWPs.

\begin{equation}\label{u_non_ideal2}
    u_{m} = m_{22}{q} + m_{23}{u}
          = \frac{q}{2}(1 + cos\delta) +\frac{u}{2}(1-cos\delta)  %-\frac{v}{\sqrt{2}} sin(\delta)
\end{equation}

\begin{figure}
\centering
\includegraphics[width=1\linewidth]{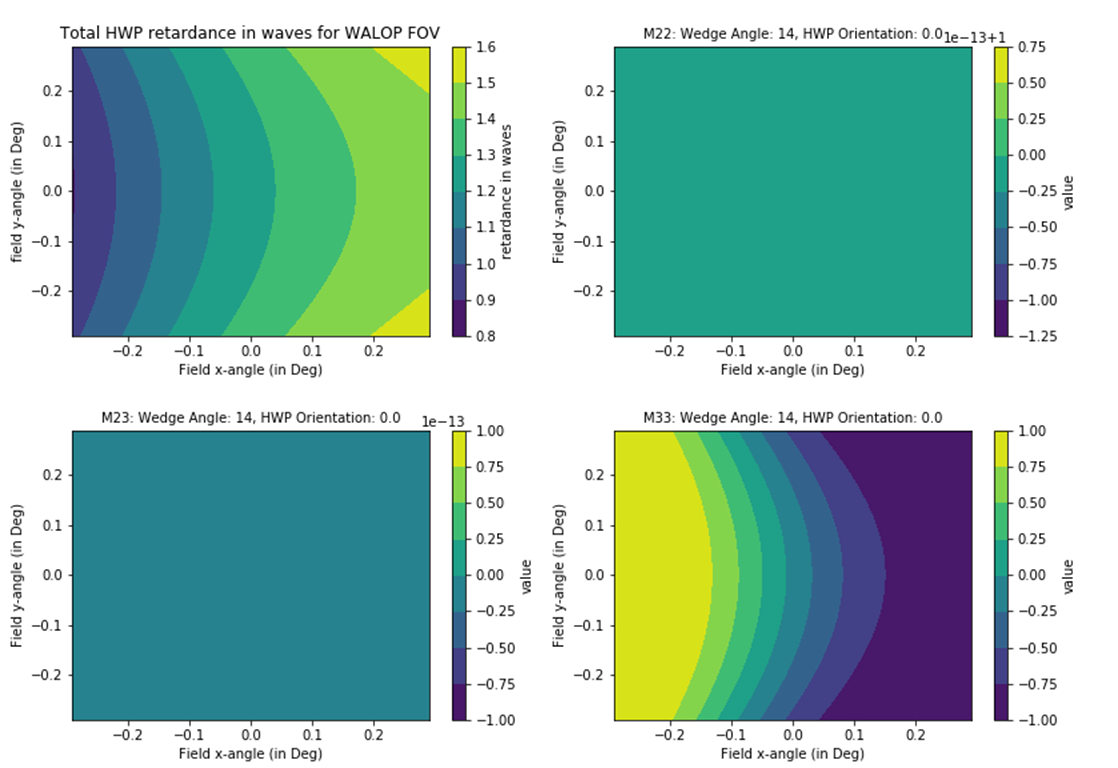}
\caption{Polarimetric behaviour of HWP1 of the Wollaston Prism Assembly (with its fast axis aligned at $0^{\circ}$ with the instrument coordinate system). While the retardance is not $\lambda/2$ across the entire FoV, the polarimetric efficiency ($M_{22}$) is 1 and cross-talk ($M_{23}$) is 0 across the FoV.}%, The polarimetric efficiency ($M_{23}$), ie, $u$ to $q_{m}$ conversion is 1 in the central areas and tends to decrease going outwards radially. Correspondingly, $M_{22}$ and $M_{33}$ are nearly zero at the center and increases in areas where the $M_{23}$ decreases}
\label{walop_LHWP}
\end{figure}

\begin{figure}
\centering
\includegraphics[width=1\linewidth]{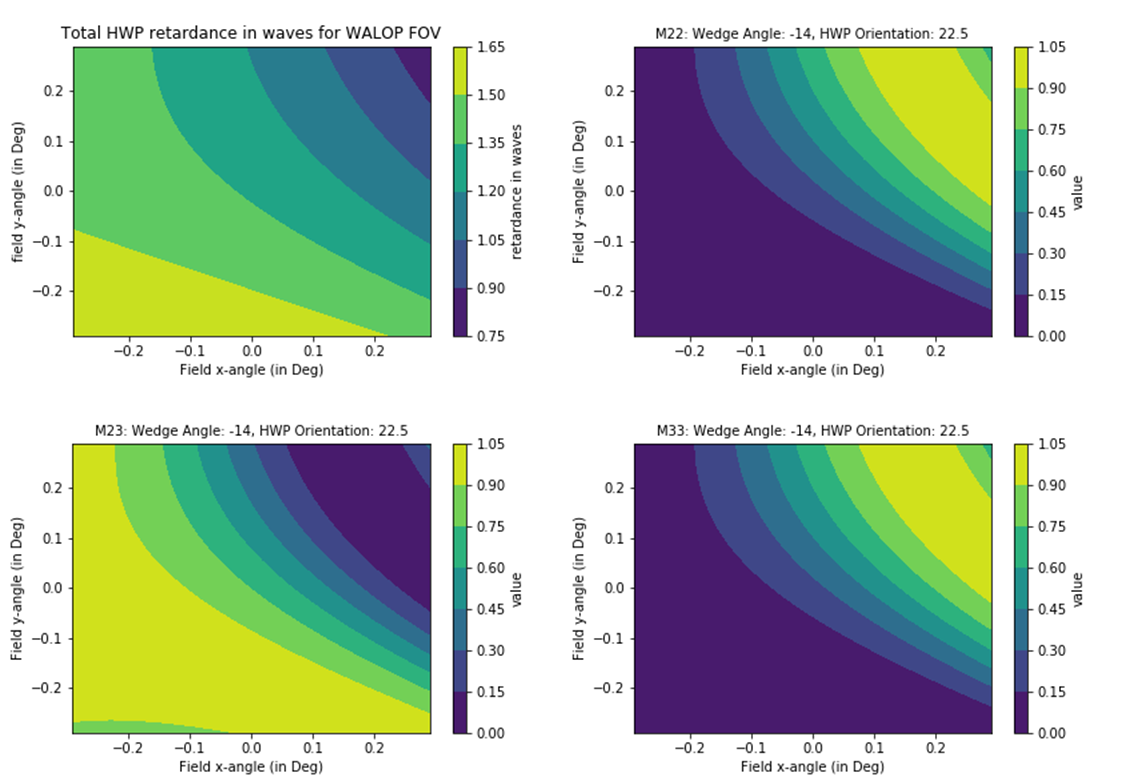}
\caption{Polarimetric behaviour of HWP2 of the Wollaston Prism Assembly (fast axis aligned at $22.5^{\circ}$ with the ICS). The retardance is not $\lambda/2$ across the entire FoV; the polarimetric efficiency ($M_{23}$) and cross-talk ($M_{23}$) varies significantly. In some places, $M_{23}$ tends to 0 and $M_{23}$ tends to 1.}
\label{walop_RHWP}
\end{figure}

The WALOP HWPs are made of one quartz and one MgF$_{2}$ plate aligned orthogonally to each other and the overall fast-axis of the HWP is along the fast-axis of the MgF$_{2}$ plate (the details of the HWP can be provided upon request).  The relative fast axis orientation of the quartz and MgF2 plates is same in HWP2 as HWP1, but are rotated by $22.5^{\circ}$ with respect to the x-axis. The prescription for finding the retardance of a wave-plate for any given incident and azimuth angle is described in the paper by Gu et al. \cite{HWP_retardance}, and elaborated below. We used their method in conjugation with ray tracing equations to model the retardance and the Mueller matrices for HWP1 and HWP2 for the entire WALOP-South FoV. The generalized Mueller matrix M for a retarder given by Equation~\ref{genera_retarder_matrix} is always in normalized form, i.e, $M_{\alpha, \delta}$ = $m_{\alpha, \delta}$. Fig~\ref{walop_LHWP} shows the retardance, $M_{22}$, $M_{33}$ and $M_{23}$ maps for HWP1 and Fig~\ref{walop_RHWP} shows the corresponding maps for HWP2. As can be seen, the patterns of polarimetric cross-talk and efficiency in the HWPs is identical to that obtained for the entire instrument in Section~\ref{WALOP_modelling_section_overall}. For example, if we look at the cross-talk term for $u_{m}$ ($M_{22}$ in Fig~\ref{walop_RHWP}), it is identical to the cross-talk term for $u_{m}$  for the entire instrument in Fig~\ref{WALOP-South Instrument Polarization}.% So we can conclude that all the cross-talk and polarimetric inefficiency for the Stokes parameters comes from the HWPs in the WPA. %Two HWPs are used in the Wollaston Prism Assembly, called as HWP1 (oriented at $0^{\circ}$ with respect to the Instrument Coordinate System (ICS)) and HWP2 (oriented at $22.5^{\circ}$ with respect to the ICS). We show the results at $0.66~{\mu}m$, representative of the behaviour of for the entire SDSS-r  filter. Fig~\ref{Mueller Matrix for HWP at 0 deg.} shows the retardance, $M_{22}$, $M_{33}$ and $M_{32}$ maps for the WALOP-South FoV for HWP1 if there were no BK-7 wedge in front of it- i.e., the central point has normal incidence. Fig~\ref{Mueller Matrix for HWP at 22.5 deg.} is the corresponding plot for the HWP2. These will be the retardance and Mueller matrix plots for the calibration HWP used in the instrument as it is also placed near the pupil.

\begin{figure}
    \centering
    \includegraphics[scale = 0.5]{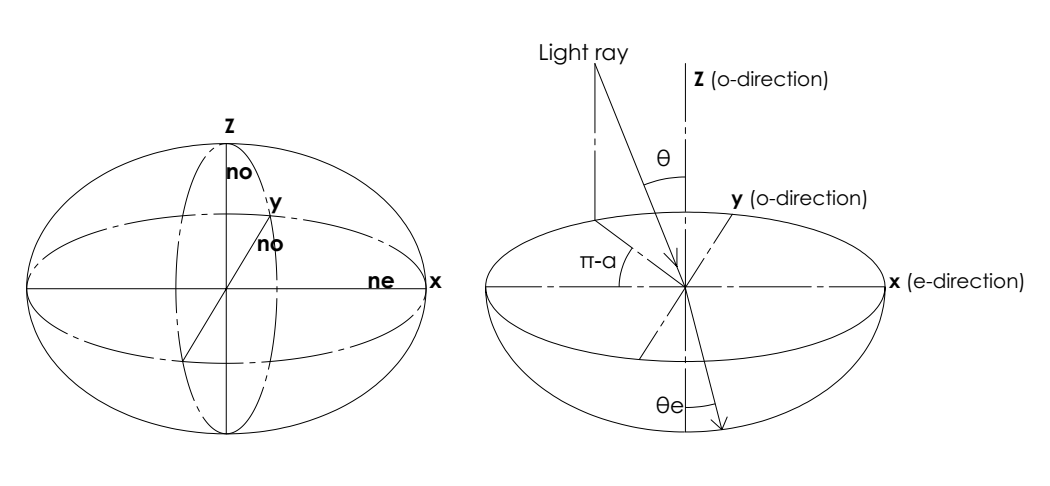}
    \caption{Left: refractive indices of uniaxial materials like $MgF_{2}$ and quartz form an ellipsoid. Right: The propagation of a light beam at an oblique angle depends on the incidence angle $\theta$ and the azimuth angle $\alpha$.}
    \label{oblique_angle_ray_propagation}
\end{figure}

When a ray enters a birefringent medium at an oblique angle, the effective refractive indices experienced by the ray is a combination of both the ordinary and extraordinary indices which depends on the azimuth as well as the incident angle, $\alpha$ and $\theta$, as shown in Fig~\ref{oblique_angle_ray_propagation}. The general optical path difference (OPD), $L_k$ and retardance $\delta_{k}$ for a retarder plate $k$ with thickness $d_k$ and retardance $\Delta{n}$ are given by the Equations~\ref{opd} and \ref{ret}.%A very good reference on this subject is the work by Gu et al. \cite{HWP_retardance, collet}. 

\begin{equation}\label{opd}
    L_{k} = \Delta{n}\times{d_{k}}
\end{equation}

\begin{equation}\label{ret}
    \delta_{k} =  \frac{2\pi}{\lambda}L_{k}
\end{equation}
% Find birefringence for  MgF2/quartz waveplate.

For a HWP made of two wave plates, the net retardance comes out to be:

\begin{equation}
    \delta = \delta_{quartz} + \delta_{MgF_{2}} = \frac{2\pi}{\lambda}\{L_{quartz} + L_{MgF_{2}}\}
\end{equation}

%def ne_eff_x(ne, no, theta, alpha):
For a wave plate aligned along the x-axis (in this case the $MgF_{2}$ plate), while the ordinary refractive index experience by the light ray remains same irrespective of $\alpha$ and $\theta$, the effective extraordinary refractive index, $ne^{'}_{x}$ is given by Equation~\ref{ne_eff}. Consequently, $n_{x} = ne^{'}_{x}$ and $n_{y} =  no$. Carrying out further calculations (presented in Gu et al.) yields the general retardance formula for a wave plate for any given $\alpha$ and $\theta$, as shown by Equation~\ref{general_ret}.

\begin{equation}\label{ne_eff}
ne^{'}_{x} = ne\sqrt{1 + (\frac{1}{n_{e}^{2}} - \frac{1}{n_{o}^2}){sin^{2}\theta}{cos^{2}\alpha}}
\end{equation}

\begin{equation}\label{general_ret}
\delta_{k} = \frac{2\pi}{\lambda}d_{k}({\sqrt{{n_{xk}}^{2} - sin^{2}\theta} - \sqrt{{n_{yk}}^{2} - sin^{2}\theta}} )
\end{equation}

For the quartz wave plate aligned along the y-axis, similar calculations can be carried out by considering the effective rotation of $90^{\circ}$ and accounting for the change in azimuth angles. Following the above steps, the retardance and subsequent Mueller matrix for HWP1 can be found. Similar calculations can be done for HWP2, by considering the overall rotation of both the quartz and MgF$_{2}$ plates as a shift in the azimuth angles by $22.5^{\circ}$. Python programs were written to carry out these calculations for the complete WALOP-South FoV and the resultant plots are shown in Figs~\ref{walop_LHWP} and \ref{walop_RHWP}.

\section{Wavelength dependence of Instrumental Polarization}\label{specpol_appendix}
\begin{figure}
    \centering
    \includegraphics[scale =0.4,trim={0 0 0 0}]{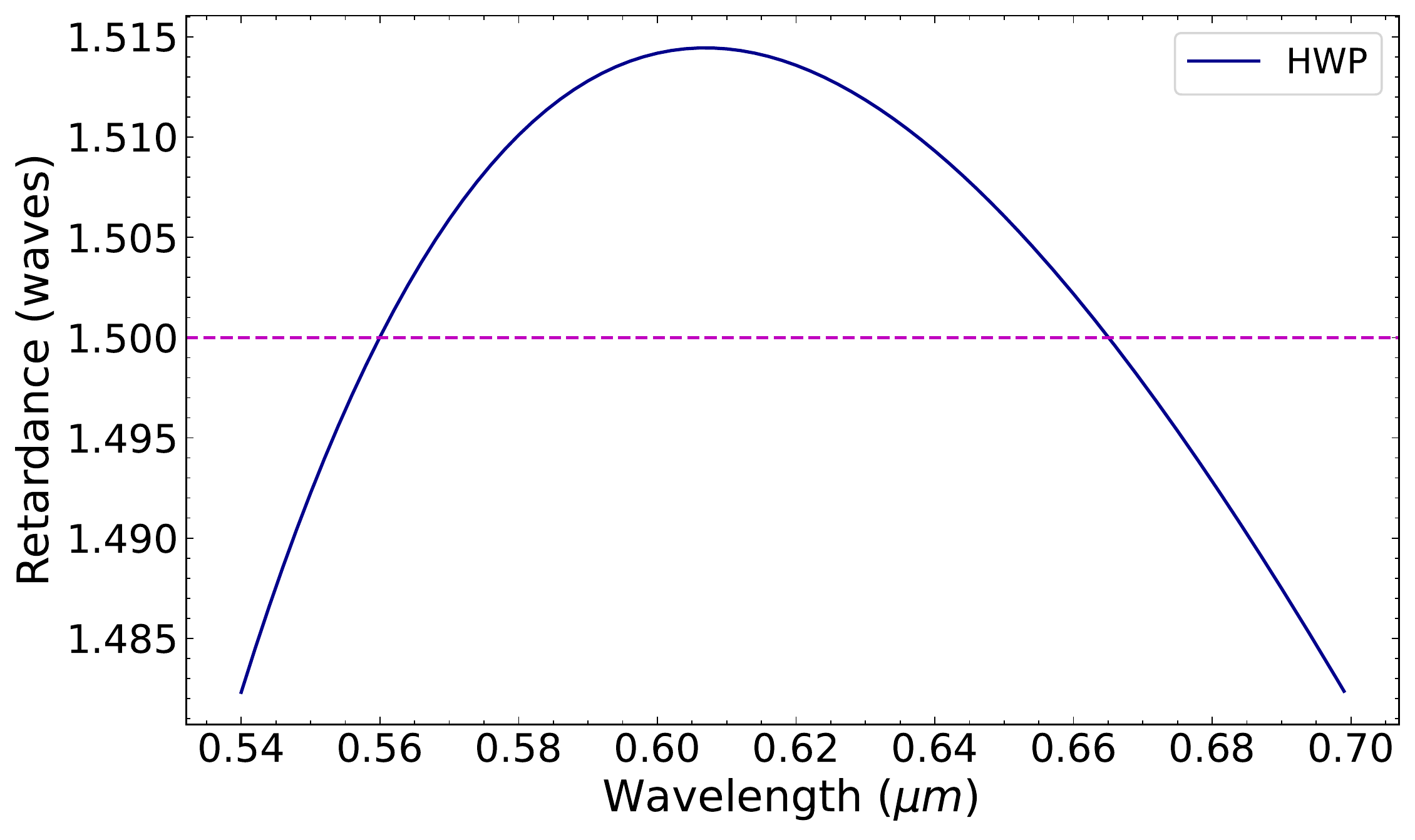}
    \caption{Retardance of the HWPs used in the WALOP-South instrument as a function of wavelength across the SDSS-r broadband filter .}
    \label{hwp_ret_wavelength}
\end{figure}
\begin{figure}
\centering
\includegraphics[width=1\linewidth]{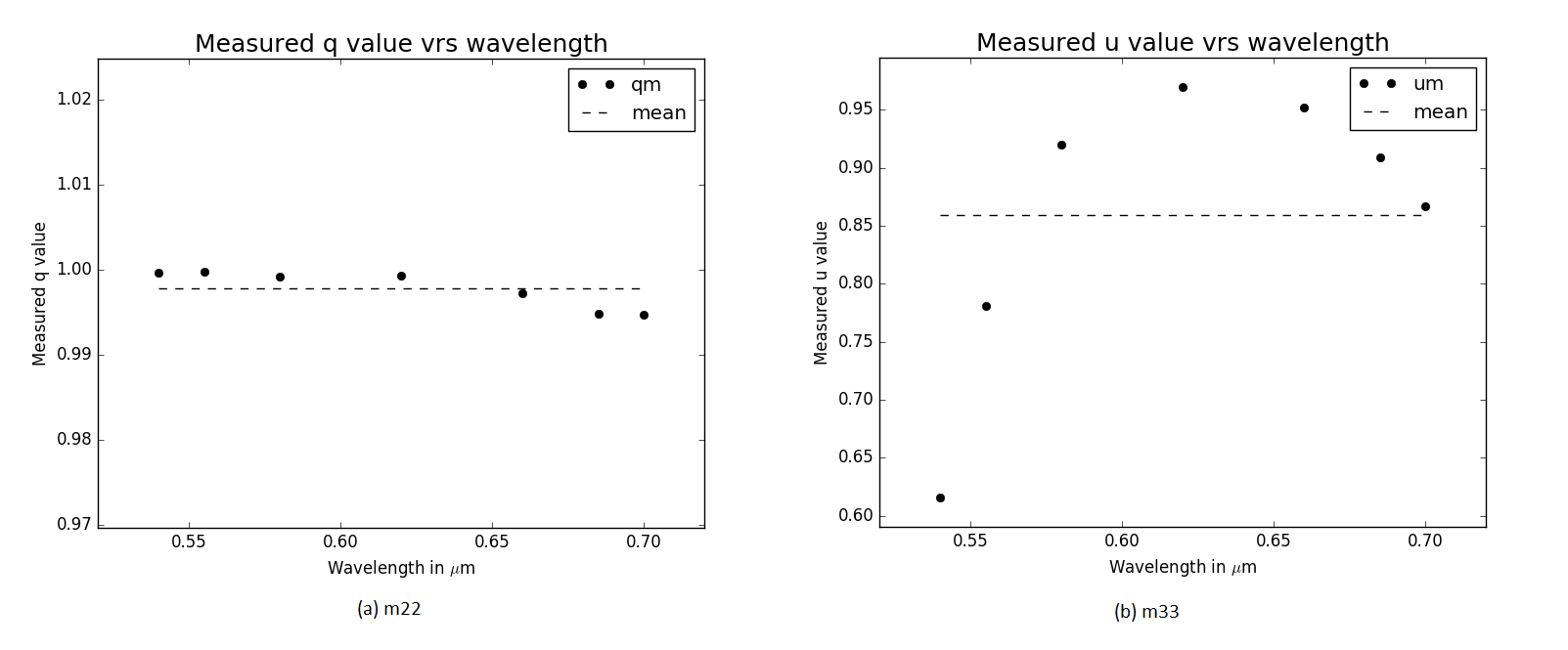}
\caption{Polarimetric efficiency of the instrument as a function of wavelength across the SDSS-r filter for the extreme field point: grid point 1.}
\label{specpol}
\end{figure}
\begin{figure}
\includegraphics[width=1\linewidth]{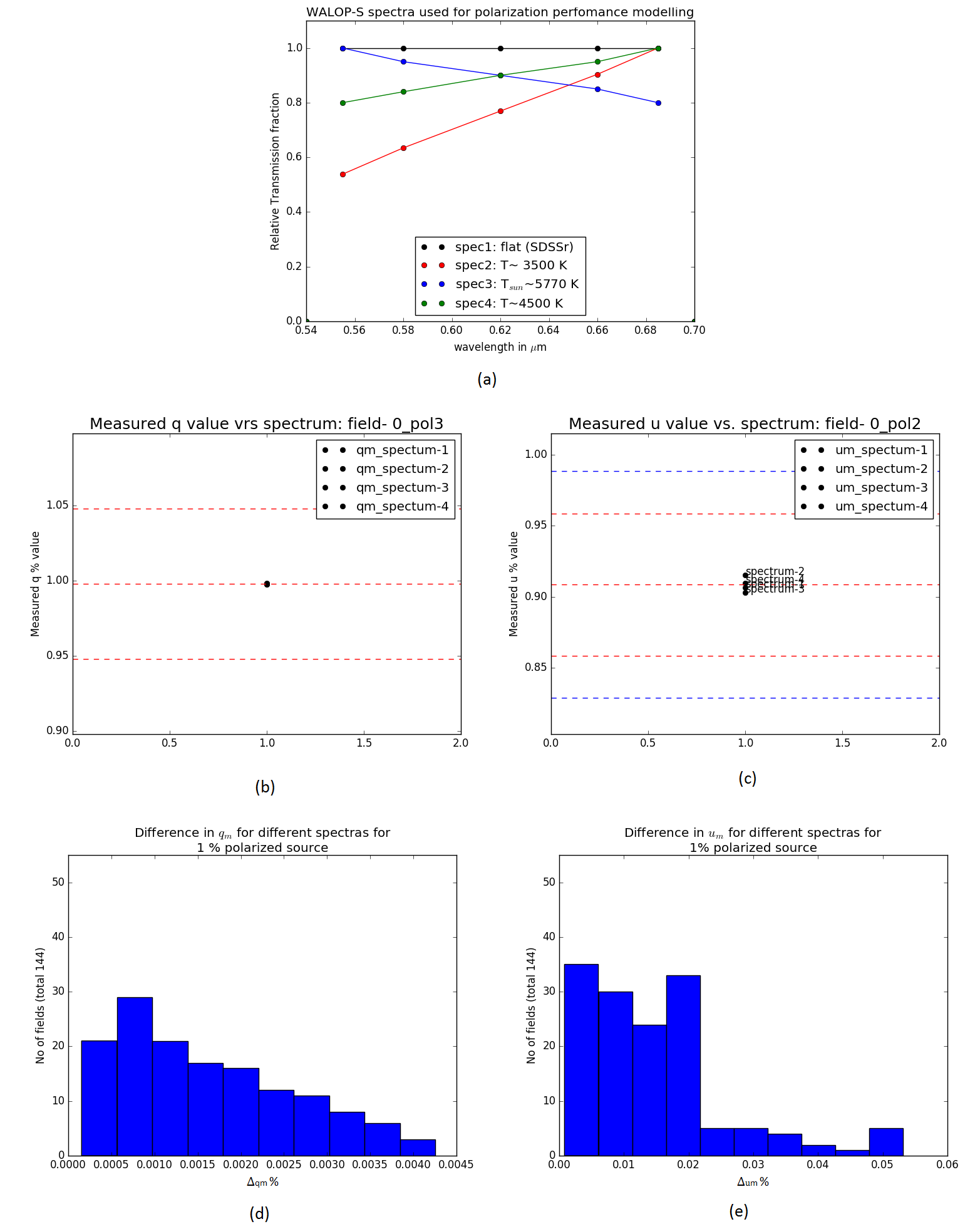}
\caption{Instrument's polarization behavior as a function of different incident spectra. (a) the four spectra used for this work, (b) and  (c) show the measured values $q_m$ and $u_m$ for a 1~\% polarized source polarization, (d) and (e) plot the histogram of the maximum difference in $q_m$ and $u_m$ for the four spectra for the 144 grid points, respectively. }
\label{instrument_specpol_model}
\end{figure}

Fig~\ref{hwp_ret_wavelength} plots the retardance of the HWPs used in the instrument across the SDSS-r filter for normal incidence. As described above, the HWP retardance varies across the WALOP-South FoV due to the angle of incidence. Consequently, the instrument's polarimetric behavior (cross-talk) varies as a function of wavelength within the SDSS-r band and the pattern of this variation is a function of the field position. Fig~\ref{specpol} plots the \textit{polarimetric efficiencies} across the SDSS-r band, i.e., measured $q$ and $u$  for input polarizations of $q$ =1 and $u$ = 1, respectively, for grid point 1 for a flat spectrum in SDSS-r filter. To estimate the effect of different spectra on the measured Stokes parameters, we fed as input stellar spectra corresponding to different effective temperatures of a star to the Zemax optical model, as shown in Fig~\ref{instrument_specpol_model}~(a). As noted earlier, the majority of the stars in the high Galactic latitude areas targeted by the \textsc{pasiphae} survey are expected to have $p < 0.3~\%$\cite{Skalidis}. We calculated the measured $q$ and $u$ values for $p$ = 1~\% and different EVPAs. Figs~\ref{instrument_specpol_model}~(b) and (c) show the measured $q$ and $u$ values for different spectra for $q$ =1 and $u$ = 1, respectively, for grid point 1. The variation is negligible for $q_m$ while it is of the order of few hundredths of a percent for $u_m$. This variation scales with $p$ of the source and will be further smaller for $p< 0.3~\%$. These variations remain similar for any EVPA for a given $p$. The deviation is more prominent for low-temperature stars ($T< 4000~K$) as their spectral transmission has a steep curve in the SDSS-r filter. Fig~\ref{instrument_specpol_model}~(d) and (e) are plotted for the extent of spread in $q_m$ and $u_m$ for the four spectral types across the 144 grid points of the FoV; $\Delta_{qm}$ and $\Delta_{um}$ are the difference between the maximum and minimum values among the spectra for the corresponding quantities. The effect is negligible for $q_m$ across the FoV. For $u_m$, there is non-negligible dependence of instrument measured $u$ on the spectral type of the star. Yet, the difference is much less than 0.05~\% for almost all field points, which is the target sensitivity of the instrument.
\par The true estimate of the instrument's on-sky wavelength-dependent polarimetric behavior will be carried out during the commissioning. For this, we have made provision for placing multiple narrowband filters in the filter wheel to facilitate modeling the instrument's polarization behavior across the SDSS-r filter. To correct for any observed dependence on the spectral class of the star, the calibration model can be developed with the spectral class/temperature of the star as a parameter, i.e., $q_i/u_i = f(q_m, u_m, s)$, where s is the spectral type of the star. \textsc{GAIA} data release 3\cite{GAIA_DR3_paper} will provide spectroscopic classification and effective temperature of all stars in the \textsc{pasiphae} survey footprint area. This information will be fed into the calibration model development.

\iffalse
\section{WALOP Calibration Model- Lab Test Results }\label{lab_calibration_appendix}

Figs~\ref{calibration_results_iqu2} to \ref{calibration_results_iqu5} are plots of the calibration model performance for cross-talk values (corresponding to different tilt angles of the HWP) from Position No 2 to 5 in Table~\ref{Calibration_lab_Results}.

\begin{figure}[htbp!]
\centering
\frame{\includegraphics[scale = 0.35]{iqu_modelling_results_tilt_2.png}}
\caption{Calibration test results for HWP tilt position 2.}
\label{calibration_results_iqu2}
\end{figure}

\begin{figure}
\centering
\frame{\includegraphics[scale = 0.35]{iqu_modelling_results_tilt_3.png}}
\caption{Calibration test results for HWP tilt position 3.}
\label{calibration_results_iqu3}
\end{figure}

\begin{figure}
\centering
\frame{\includegraphics[scale = 0.35]{iqu_modelling_results_tilt_4.png}}
\caption{Calibration test results for HWP tilt position 4.}
\label{calibration_results_iqu4}
\end{figure}

\begin{figure}
\centering
\frame{\includegraphics[scale = 0.35]{iqu_modelling_results_tilt_5.png}}
\caption{Calibration test results for HWP tilt position 5.}
\label{calibration_results_iqu5}
\end{figure}
\fi

\acknowledgments 
\par The \textsc{pasiphae} program is supported by grants from the European Research Council (ERC) under
grant agreements No 771282 and No 772253; by the National Science Foundation (NSF) 
award AST-2109127;  by the National Research Foundation of South Africa under the National
Equipment Programme; by the Stavros Niarchos Foundation under grant \textsc{pasiphae}  ; and by the Infosys Foundation. 

This work utilized the open source software packages Astropy\cite{astropy:2013, astropy:2018}, Numpy\cite{numpy}, Scipy\cite{scipy}, Matplotlib\cite{matplotlib} and Jupyter notebook\cite{jupyter_notebook}.

\par We are thankful to Vinod Vats at Karl Lambrecht Corp. for his inputs and suggestions on various aspects of half-wave plate design and performance. 

%%%%% References %%%%%

\bibliography{article}   % bibliography data in report.bib
\bibliographystyle{spiejour}   % makes bibtex use spiejour.bst
\appendix

\end{spacing}
\end{document}